\RequirePackage{fix-cm}
\documentclass[twocolumn]{svjour3}          
\smartqed  
\usepackage{graphicx}
\usepackage[utf8]{inputenc}
\usepackage[T1]{fontenc}
\usepackage[english]{babel} 
\usepackage{amsmath}
\usepackage{amsfonts}
\usepackage{amssymb}
\usepackage{makeidx}
\usepackage{graphicx}
\usepackage{color} 
\usepackage[dvipsnames]{xcolor}
\usepackage{enumerate} 
\usepackage{enumitem} 
\usepackage{lmodern} 
\usepackage{wasysym} 
\usepackage{booktabs} 
\usepackage{float} 
\usepackage{subfigure}
\usepackage{lipsum} 
\usepackage{textcomp}
\usepackage{chngcntr}
\usepackage{xcolor} 
\usepackage{multirow}
\usepackage{todonotes}
\usepackage{tikz}
\usepackage{pgfplots}
\pgfplotsset{compat=newest}
\pgfplotsset{plot coordinates/math parser=false}
\pgfplotsset{ymissing/.style={ylabel={}, yticklabels={}}}
\pgfplotsset{xmissing/.style={xlabel={}, xticklabels={}}}
\usepgfplotslibrary{external} 
\usetikzlibrary{spy,shapes,decorations.pathmorphing,plotmarks,arrows,automata,patterns,decorations.text,calc,backgrounds,shadows,positioning,angles,quotes,arrows.meta,external}
\newcommand{\RN}[1]{%
	\textup{\uppercase\expandafter{\romannumeral#1}}%
}
 \journalname{Preprint}
\begin{document}

\title{How to design a 2D active grid for dynamic inflow modulation
}


\author{Tom T. B. Wester$^1$ \and 
	Johannes Krauss$^1$ \and 
	 Lars Neuhaus$^1$ \and 
	  Agnieszka Hölling$^1$ \and 
	   Gerd Gülker$^1$ \and 
	    Michael Hölling$^1$ \and 
	     Joachim Peinke$^1$   
}


\institute{$^1$ForWind, Institute of Physics, University of Oldenburg\\
              Küpkersweg 70 \\
              Tel.: +49-441-7985023\\
              Fax: +49-441-7985099\\
              \email{tom.wester@uni-oldenburg.de}           
}

\maketitle

\begin{abstract}
Wind turbines operate under constantly changing turbulent inflow conditions. In the rotating system, wind gusts lead to variations in the angle of attack at local blade segments resulting in dynamic effects such as dynamic stall. Such highly non-linear effects are known to produce a significant overshoot in the lift and thus an increase in loads acting on the wind turbine, leading to long-term fatigue. To better understand these effects, it is essential to perform experiments under defined conditions on 2D airfoil segments in the wind tunnel.\\
In this study, a so-called 2D active grid is presented which allows to generate local inflow conditions with defined fluctuations of the angle of incidence (AoI) in wind tunnel experiments. The focus of the investigations is on sinusoidal variations of AoI with high amplitudes generated by different grid configurations. By changing the AoI dynamic phenomena can be induced without the need to move the object under investigation. Inertial effects during force measurements and a changing shadow casting due to a moving airfoil in particle image velocimetry measurements do not appear. Additional variations in the longitudinal velocity component are another aspect in the presented work. Such longitudinal gusts can be combined with AoI variations in arbitrary phase. This can be used to mimic various inflow situations such as yaw or tower shadow effects on wind turbines. 
\end{abstract}

\section{Introduction}
\label{chap:Introduction}
\begin{figure}[htpb]
	\centering
	\includegraphics[width=.99\linewidth]{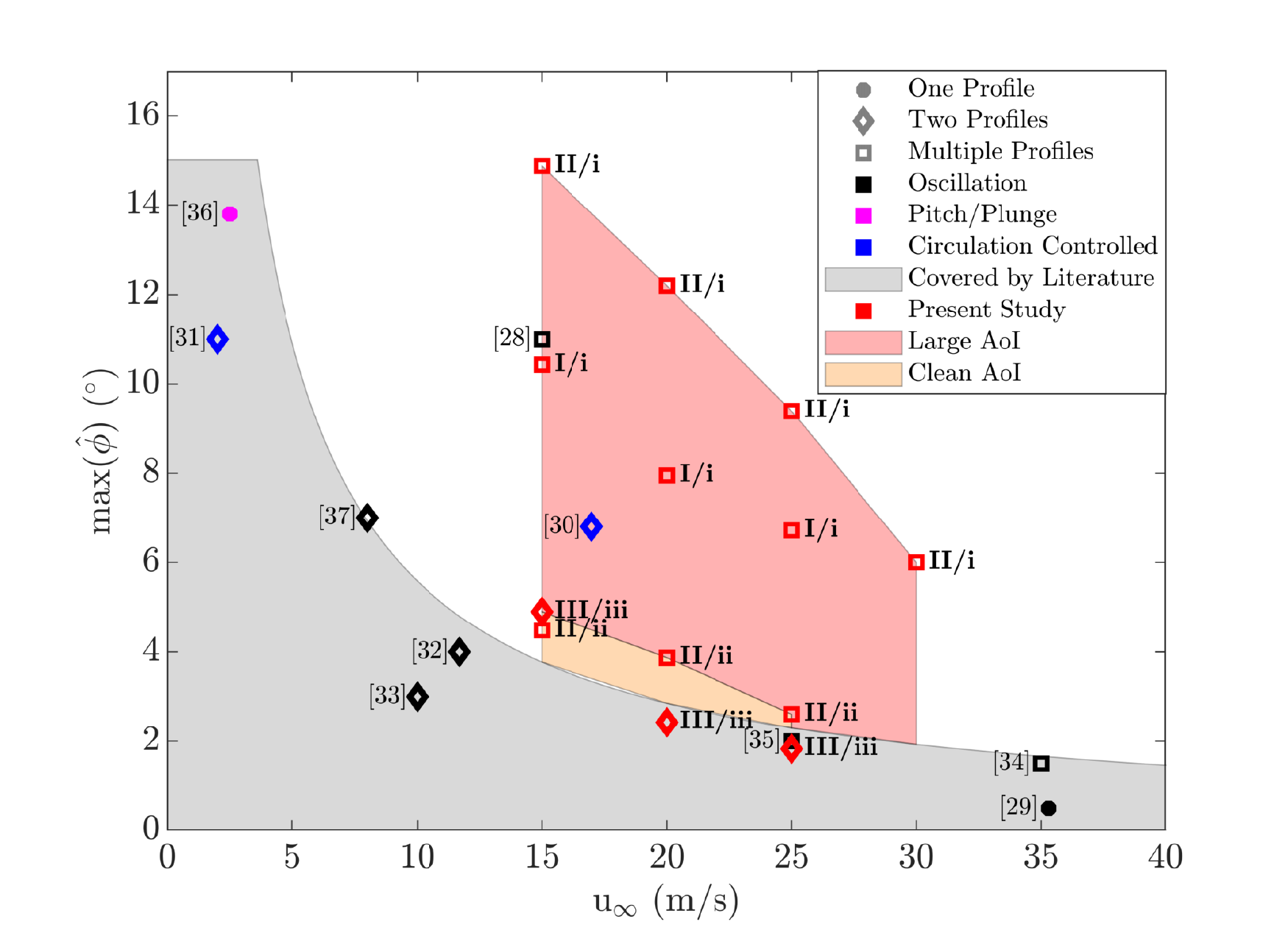}
	\caption[Literature comparison]{Plot of maximal achieved vertical gust amplitudes in the literature as a function of velocity based on following studies: \cite{Grissom2004,Hakkinen1957,Ham1974,Jancauskas1986,Stapountzis1982,Wu2013,Kobayakawa1978,Tang1996,Wei2019b,Wood2017}. The symbols indicate the number of used profiles (one ($\circ$), two($\diamond$), multiple($\square$)). The approach is indicated by the color (oscillation (\textcolor{black}{$\blacksquare$}), pitch/plunge(\textcolor{magenta}{$\blacksquare$}), circulation controlled (\textcolor{blue}{$\blacksquare$})). The gray area indicates what has been shown in closed test sections with oscillating profiles so far in literature. The results of this study are plotted in red, and the newly covered area for very clean AoI modulations in orange and very large AoI variations in red.}
	\label{fig:Lit_Comp}
\end{figure}

Under rapidly changing inflow conditions, the aerodynamics of objects becomes challenging and highly non-linear. Exemplary of such objects are wind turbines operating in the atmospheric boundary layer. Under frequently occurring so-called extreme conditions, which manifest as gusts or sudden changes in direction of the incoming flow, the turbines are exposed to rapid inflow fluctuations. For the blade in the rotating system, these changes represent a variation in the angle of incidence (AoI), which leads to effects such as dynamic stall. This causes dynamic loads acting on the blade, which are passed on from there to the drive train and the power electronics of the wind turbine, leading to an increase in fatigue and thus to a reduction in the service life of the entire turbine \cite{spinato2009reliability}. The amplitude of such AoI changes or vertical gusts can easily exceed several degrees, depending on the operating point of the turbine and span position.\\
The aerodynamic phenomenon of dynamic stall was first observed nearly a century ago during two dimensional measurements of a pitching Göttingen 459 profile \cite{kramer1932zunahme}. Since then, the phenomenon has received more attention in the field of helicopter aerodynamics \cite{harris1968blade,liiva1969unsteady}, where the dynamic stall is caused by the rapidly changing relative speed of the advancing or retrieving rotor of the helicopter and the additional periodic pitch movement. Up to now most dynamic stall experiments and CFD simulations are performed using pitching airfoils \cite{mccroskey1972dynamic,mulleners2013dynamic,Choudhry2014,Dunne2015}. For these experiments, a pitch movement of more than 5° is usually selected to generate a sufficiently large dynamic stall cycle. \\
For wind turbine rotors, the dynamic stall is mainly caused by the changing inflow rather than a pitching of the blades. Although studies have shown the effects occurring to be comparable \cite{gharali2012numerical,rival2010characteristics}, the situation at hand is different. Whereas pitching airfoils change the angle of attack along the entire chord at the same time, a varying inflow provides different angles of attack along the chord. The faster these changes take place, the more significant this effect is. For applications where the inflow and not the object's angle of attack changes, this should be handled the same way during the experiment. This creates the need to be able to generate large AoI fluctuations in the wind tunnel. \\
The generated amplitudes must be comparable to the pitching airfoil experiments. This is the only possibility to perform realistic dynamic stall experiments with respect to wind turbines in a 2D model. Additionally, this configuration simplifies particle image velocimetry (PIV) and force measurements on the airfoil, as the force signals are not subject to any additional inertia. The light sheet for PIV measurements does not need to be adjusted individually but can remain unchanged for different AoI amplitudes at fixed geometric angle of attack (AoA) $\alpha$ of the investigated airfoil.\\
Besides the investigation of non-linear aerodynamic effects, the validation of fundamental theoretical considerations is also a motivation for the generation of vertical gusts. Sears already modeled the influence of an incoming vertical gust on airfoils with the thin-airfoil theory in the early 20th century \cite{sears1938}. He described the vertical gust as a distribution of vortices along the airfoil camber. This theory was later expanded by Goldstein and Atassi \cite{goldstein1976complete} and Atassi himself \cite{atassi1984} with second order models.\\
Proving these theories experimentally turned out to be very difficult. The use of a classical 3D active grid \cite{Makita1991,knebel2011,Reinke_PHD_2017} for flow modulation was the first way to achieve agreement between theory and experiment \cite{cordes2017}. Subsequent studies on the 3D grid have revealed a three-dimensionality of the generated flow \cite{traphan2020dynamic}. This three-dimensionality is probably responsible for the ambiguity between the theories of Sears and Atassi in \cite{cordes2017}. The ambiguity could be eliminated by using the 2D active grid presented here \cite{Wei_2019}. The results emphasize the need for a device enabling both Sears conditions (no fluctuations in the $u_x$ component) and Atassi conditions (additional modulation of $u_x$ component in the gust).\\
The idea of generating periodic two-dimensional longitudinal and vertical gusts in wind tunnels has now existed for over half a century. A plot of achieved AoI amplitudes in different studies using closed test sections is shown in Fig. \ref{fig:Lit_Comp}. Here, the maximum amplitude of the vertical gust $\hat{\phi}$ achieved in each study is plotted against the velocity u$_\infty$ of the wind tunnel. The studies so far show an uncovered range for larger velocities and amplitudes. This area could only be reached by studies using an open jet wind tunnel (\cite{Ricci2008,lancelot2017design,Kuzmina2006,Saddington2015}). Since the fluid dynamic fundamentals are different due to the open test section, those studies were excluded from the plot. In addition to the studies shown here for the low velocity range, the successful application of oscillating profiles could also be shown in transonic wind tunnels (\cite{GILMAN1966,ReedIII1981,Neumann2013,Brion_2015}). Here, however, the amplitudes are reported to be below 1.5$^\circ$.
\\
Since small velocities are seldom important for aerodynamic experiments and large AoI amplitudes are needed for comparison with typical dynamic stall experiments, this study presents a device which can reach the area not yet covered.\\
Besides desired AoI variations, the presented 2D grid is also capable of inducing longitudinal gusts with large amplitudes at small-time scales. Such velocity gusts can be generated using different approaches. In \cite{neunaber2020aerodynamic} a new device called "chopper" was introduced, which generates velocity fluctuations up to 0.4Hz upstream of the test section. In \cite{greenblatt2016}, an overview of louver systems for modulating flow velocity is given, but no modulation with more than 4.2Hz is mentioned. Recently \cite{Farnsworth2020} introduced the idea of using a louver system in front of the wind tunnel fan to modulate the inflow velocity with up to 1.5Hz. The louver systems show clear advantages in the quality of the generated gust, but they are also limited to this application. For the 2D active grid the separability of the individual shaft motion allows one part to be used for AoI and another for velocity fluctuations, enabling a combination of longitudinal and vertical gust with only one device.\\
The paper is structured as follows. First the used experimental setup is presented in Sec. \ref{chap:ExperimentalSetup}. This includes the wind tunnel in Sec. \ref{chap:Windtunnel} and the new 2D active grid in Sec. \ref{chap:2DGrid} with its different configurations and shaft shapes. In the following, different possible inflows are discussed in Sec. \ref{chap:Inflow}. In a first step the two-dimensionality of the generated flow field is shown in Sec. \ref{chap:2DInflow}. In Sec. \ref{chap:AoImodulation} the achievable AoI variations are compared based on the grid configuration. In Sec. \ref{chap:velocitymodulation} the possible longitudinal gusts are presented, followed by the combination of AoI and velocity fluctuations in Sec. \ref{chap:bothmodulation}. An exemplary application is shown in Sec. \ref{chap:PIV}, where a PIV measurement of a profile subject to such AoI variation is shown. Finally, Sec. \ref{chap:Conclusion} concludes the paper.

\section{Experimental setup}
\label{chap:ExperimentalSetup}
\subsection{Wind tunnel}
\label{chap:Windtunnel}
\begin{figure}[htpb]
	\centering
	\includegraphics[width=.99\linewidth]{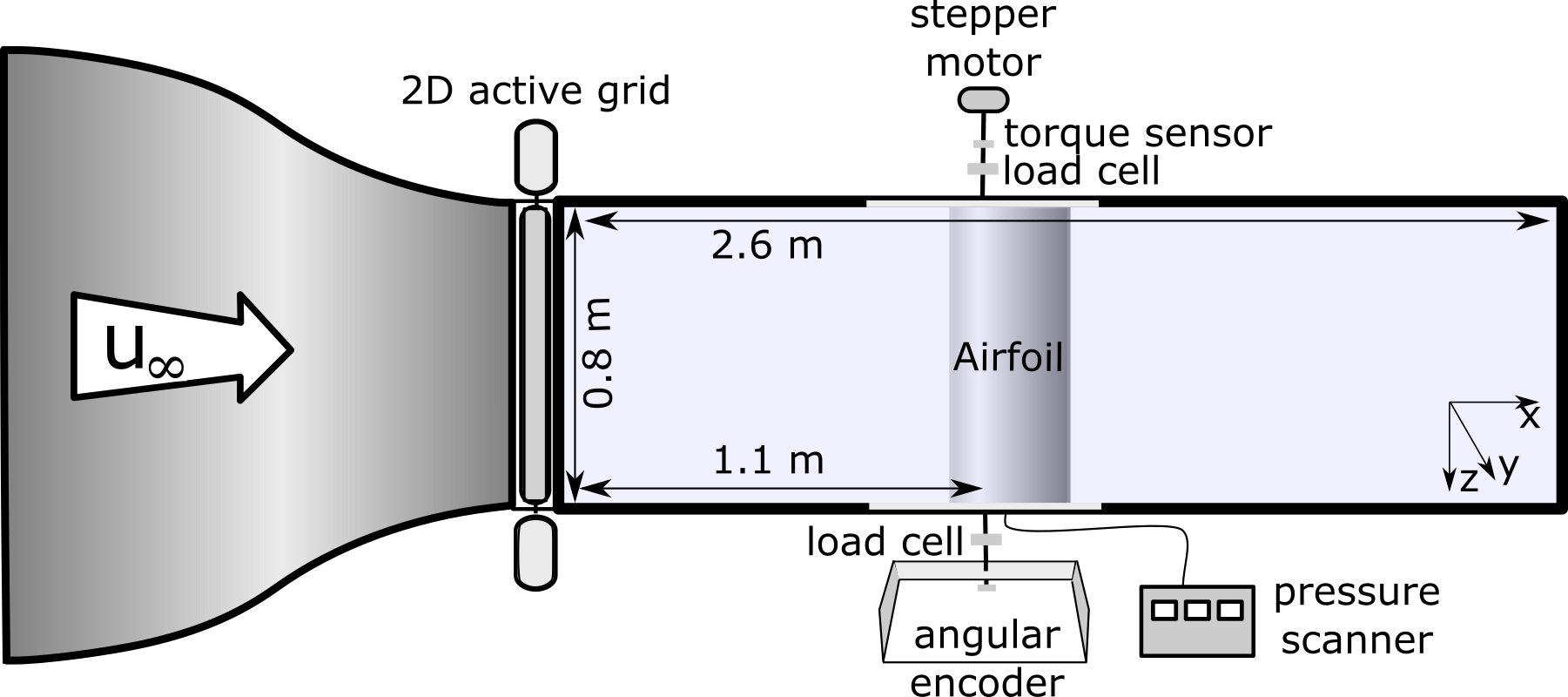}
	\caption[Wind tunnel]{Used experimental setup. Inflow is coming from left and passes the 2D active grid, which is mounted at the nozzle. The flow travels in x-direction towards the airfoil which is mounted along the z-axis in the middle of the wind tunnel.}
	\label{fig:WindtunnelSetup}
\end{figure}

The experiments are performed in a wind tunnel with a closed test section. A sketch of the tunnel is shown in Fig. \ref{fig:WindtunnelSetup}. It is a closed loop wind tunnel of Göttingen-type with a maximal speed of 50m/s. The closed test section has a cross section of (0.8$\times$1.0)m$^2$ $(h \times w)$ and a length of 2.6m. The side walls and the top and bottom of the closed test section are made of acrylic glass walls to allow full optical access for measurements with Laser Doppler Anemometer (LDA) or PIV. The figure also shows the coordinate system used for the following results.
\\
The shown wind tunnel is mainly used to measure aerodynamic forces and investigate aerodynamic effects evolving around airfoils. Such objects can be mounted $1.1$m downstream of the nozzle. The airfoils are rotated around their quarter point using a stepper motor. The geometric angle of attack $\alpha$ is recorded with an angular encoder positioned underneath the test section. Aerodynamic forces are measured using load cells with a temporal resolution of 1kHz. Those are mounted on top and bottom of the airfoil. To capture the torque, an additional sensor is placed below the stepper motor.\\
To characterize the generated flow fields X-wire measurements are performed $1$m downstream of the grid at centerline position (position of airfoil leading edge). 
The characterizations are performed using X-wires of type 55P51 from Dantec Dynamics. A Streamline 90H02 Flow Unit is used for the calibration of the X-wires. They are calibrated at 30 logarithmic distributed points in a velocity range of (1-50)m/s. For directional calibration, 17 points are chosen between -40$^\circ$ and 40$^\circ$ at given inflow velocity of the experiment. The X-wires are connected to a Streamline 90N10 frame with four CTA90C10 units. A sampling rate of 20kHz with a low pass filter of 10kHz is chosen for the measurements. The data is recorded with a DAQ MX 6281 A/D converter with a bit depth of 18 bit in a voltage range of (0-5)V. With these settings the error of the velocity measurement can be estimated with 1$\%$ \cite{Joergensen2001}. Since phase-averaged data are used during the evaluation, this error can be considered negligible.

\subsection{2D Active Grid}
\label{chap:2DGrid}
\begin{figure}[h]
	\centering
	\includegraphics[width=.75\linewidth]{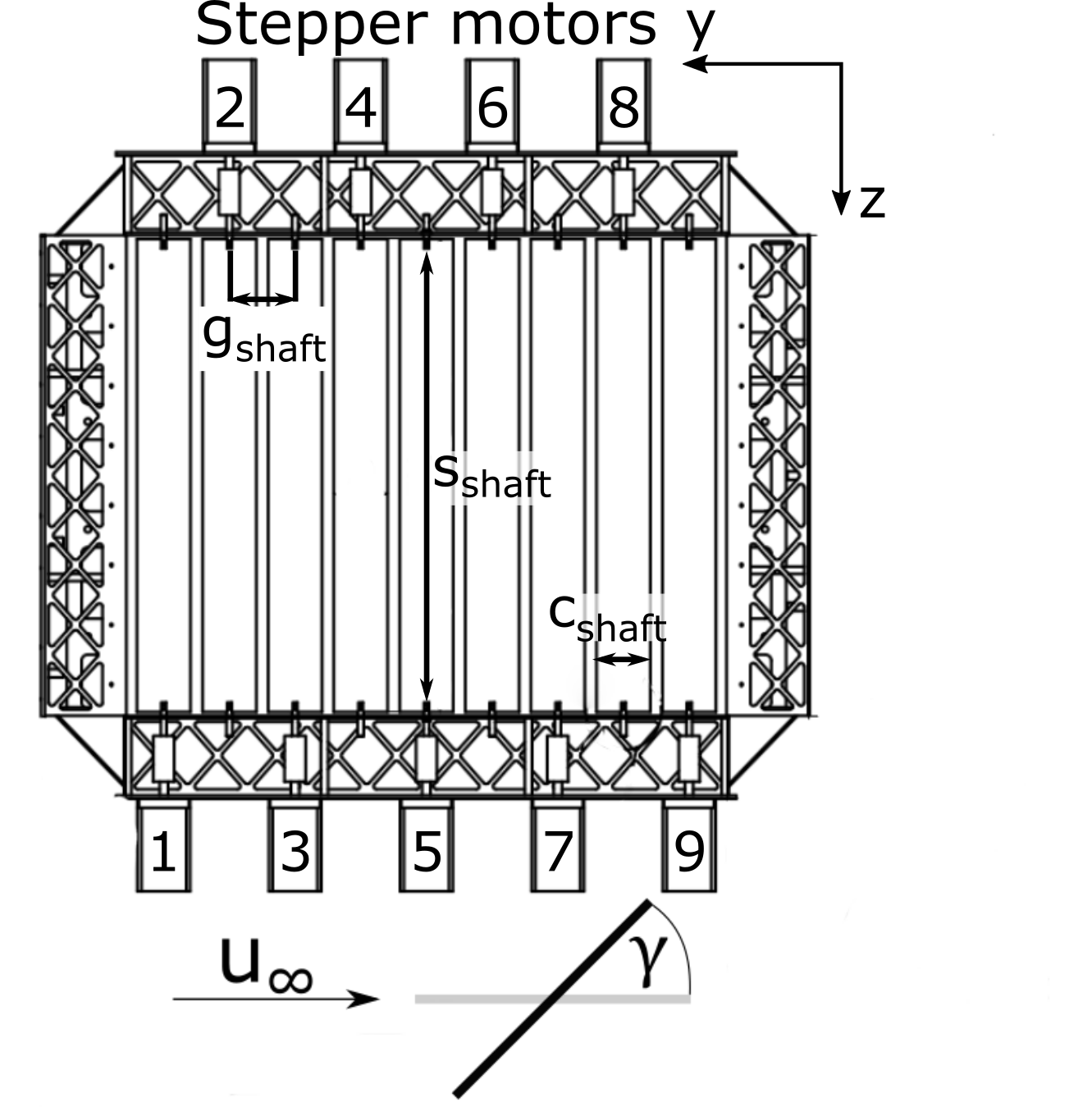}
	\caption[Reduced active grid]{Technical drawing of the 2D active grid containing the stepper motors (1-9), the gauge of the shafts $g_{shaft}$, the span of the shafts $s_{shaft}$ and the chord length of the shafts $c_{shaft}$. At the bottom the definition of shaft angle $\gamma$ is shown (top view onto the shaft).}
	\label{fig:2DGrid_Technical}
\end{figure}

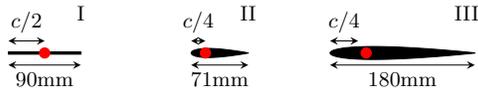
\begin{figure}[h]
	\centering
%
%
\begin{tikzpicture}

\begin{axis}[%
width=\linewidth,
axis equal image,
xmin=-60,
xmax=540,
xticklabels={},
yticklabels={},
ymin=-70,
ymax=70,
hide axis,
]

\addplot[area legend, draw=black, fill=black, forget plot]
table[row sep=crcr] {%
x	y\\
-45	-1.5\\
-45	1.5\\
45	1.5\\
45	-1.5\\
-45	-1.5\\
}--cycle;

\draw[>=stealth,<->] (-45,-15) --node[midway,below](){90mm} (45,-15);
\draw[>=stealth,<->] (-45,15) --node[midway,above](){$c/2$} (0,15);

\addplot [color=red, line width=2.5pt, draw=none, mark size=1.0pt, mark=*, mark options={solid, fill=red, red}, forget plot]
  table[row sep=crcr]{%
0	0\\
};

\node[align=right]
at (axis cs:45,50) {I};

\addplot[area legend, draw=black, fill=black, forget plot]
table[row sep=crcr] {%
x	y\\
253.25	0.11928\\
253.140589	0.139728\\
252.812924	0.200717\\
252.269135	0.301253\\
251.512488	0.43949\\
250.54774	0.613227\\
249.380713	0.819624\\
248.01872	1.055415\\
246.470068	1.317192\\
244.744413	1.601192\\
242.852263	1.90351\\
240.805404	2.220099\\
238.616403	2.547054\\
236.298679	2.880115\\
233.866645	3.215093\\
231.335282	3.547728\\
228.720068	3.873547\\
226.037333	4.187722\\
223.303407	4.485567\\
220.53533	4.76197\\
217.75	5.011677\\
214.96467	5.229434\\
212.196593	5.409987\\
209.462667	5.548366\\
206.779932	5.639743\\
204.164718	5.68\\
201.633355	5.665658\\
199.201321	5.594019\\
196.883597	5.463592\\
194.694596	5.273809\\
192.647737	5.025238\\
190.755587	4.719512\\
189.029932	4.359329\\
187.48128	3.948097\\
186.119287	3.490005\\
184.95226	2.989526\\
183.987512	2.451204\\
183.230865	1.879512\\
182.687076	1.278284\\
182.359411	0.650999\\
182.25	0\\
182.359411	-0.650999\\
182.687076	-1.278284\\
183.230865	-1.879512\\
183.987512	-2.451204\\
184.95226	-2.989526\\
186.119287	-3.490005\\
187.48128	-3.948097\\
189.029932	-4.359329\\
190.755587	-4.719512\\
192.647737	-5.025238\\
194.694596	-5.273809\\
196.883597	-5.463592\\
199.201321	-5.594019\\
201.633355	-5.665658\\
204.164718	-5.68\\
206.779932	-5.639743\\
209.462667	-5.548366\\
212.196593	-5.409987\\
214.96467	-5.229434\\
217.75	-5.011677\\
220.53533	-4.76197\\
223.303407	-4.485567\\
226.037333	-4.187722\\
228.720068	-3.873547\\
231.335282	-3.547728\\
233.866645	-3.215093\\
236.298679	-2.880115\\
238.616403	-2.547054\\
240.805404	-2.220099\\
242.852263	-1.90351\\
244.744413	-1.601192\\
246.470068	-1.317192\\
248.01872	-1.055415\\
249.380713	-0.819624\\
250.54774	-0.613227\\
251.512488	-0.43949\\
252.269135	-0.301253\\
252.812924	-0.200717\\
253.140589	-0.139728\\
253.25	-0.11928\\
}--cycle;

\draw[>=stealth,<->] (182.25,-15) --node[midway,below](){71mm} (253.25,-15);
\draw[>=stealth,<->] (182.25, 15) --node[midway,above](){$c/4$} (200,15);

\addplot [color=red, line width=2.5pt, draw=none, mark size=1.0pt, mark=*, mark options={solid, fill=red, red}, forget plot]
  table[row sep=crcr]{%
200	0\\
};

\node[align=right]
at (axis cs:253.25,50) {II};

\addplot[area legend, draw=black, fill=black, forget plot]
table[row sep=crcr] {%
x	y\\
535	0.1701\\
534.72262	0.19926\\
533.89192	0.2862\\
532.5133	0.42948\\
530.59504	0.62676\\
528.1492	0.87444\\
525.19054	1.16874\\
521.7376	1.50516\\
517.81144	1.87848\\
513.43654	2.28348\\
508.63954	2.7144\\
503.45032	3.16602\\
497.90074	3.63222\\
492.02482	4.10724\\
485.8591	4.58496\\
479.44156	5.05926\\
472.81144	5.52384\\
466.01014	5.97204\\
459.07906	6.39666\\
452.0614	6.79086\\
445	7.1469\\
437.9386	7.4574\\
430.92094	7.71498\\
423.98986	7.91226\\
417.18856	8.04258\\
410.55844	8.1\\
404.1409	8.07948\\
397.97518	7.97742\\
392.09926	7.79148\\
386.54968	7.52076\\
381.36046	7.16616\\
376.56346	6.7302\\
372.18856	6.21666\\
368.2624	5.63022\\
364.80946	4.977\\
361.8508	4.2633\\
359.40496	3.4956\\
357.4867	2.6802\\
356.10808	1.82304\\
355.27738	0.92844\\
355	0\\
355.27738	-0.92844\\
356.10808	-1.82304\\
357.4867	-2.6802\\
359.40496	-3.4956\\
361.8508	-4.2633\\
364.80946	-4.977\\
368.2624	-5.63022\\
372.18856	-6.21666\\
376.56346	-6.7302\\
381.36046	-7.16616\\
386.54968	-7.52076\\
392.09926	-7.79148\\
397.97518	-7.97742\\
404.1409	-8.07948\\
410.55844	-8.1\\
417.18856	-8.04258\\
423.98986	-7.91226\\
430.92094	-7.71498\\
437.9386	-7.4574\\
445	-7.1469\\
452.0614	-6.79086\\
459.07906	-6.39666\\
466.01014	-5.97204\\
472.81144	-5.52384\\
479.44156	-5.05926\\
485.8591	-4.58496\\
492.02482	-4.10724\\
497.90074	-3.63222\\
503.45032	-3.16602\\
508.63954	-2.7144\\
513.43654	-2.28348\\
517.81144	-1.87848\\
521.7376	-1.50516\\
525.19054	-1.16874\\
528.1492	-0.87444\\
530.59504	-0.62676\\
532.5133	-0.42948\\
533.89192	-0.2862\\
534.72262	-0.19926\\
535	-0.1701\\
}--cycle;

\draw[>=stealth,<->] (355,-15) --node[midway,below](){180mm} (535,-15);
\draw[>=stealth,<->] (355, 15) --node[midway,above](){$c/4$} (390,15);

\addplot [color=red, line width=2.5pt, draw=none, mark size=1.0pt, mark=*, mark options={solid, fill=red, red}, forget plot]
  table[row sep=crcr]{%
400	0\\
};

\node[align=right]
at (axis cs:525,50) {III};
\end{axis}

\end{tikzpicture}%
	\caption{Used shafts for the 2D active grid. \textbf{I}: flat plates with a chord length of 90mm. \textbf{II}: NACA0018 airfoils with a chord length of 71mm. \textbf{III}: NACA0009 with a chord length of 180mm. The red dot indicates the pivot point.}
	\label{fig:ShaftsTypes}
\end{figure}

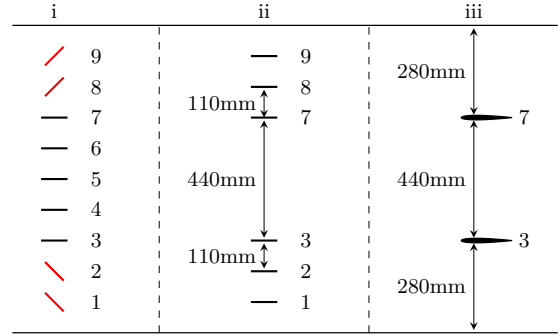
\begin{figure}[h]
	\centering
%
%
\begin{tikzpicture}

\begin{axis}[%
	width=\linewidth,
	axis equal image,
clip=false,
separate axis lines,
every outer x axis line/.append style={white!15!black},
every x tick label/.append style={font=\color{white!15!black}},
every x tick/.append style={white!15!black},
xmin=-25,
xmax=1700,
xtick={0,1000,2000},
xticklabels={{},{},{}},
every outer y axis line/.append style={white!15!black},
every y tick label/.append style={font=\color{white!15!black}},
every y tick/.append style={white!15!black},
ymin=-500,
ymax=500,
ytick={-500,0,500},
yticklabels={{},{},{}},
axis line style={draw=none},
ticks=none,
axis on top
]

\addplot[area legend, line width=0.5pt, draw=black, fill=black, forget plot]
table[row sep=crcr] {%
	x	y\\
	-150	550\\
	1800	550\\
}--cycle;

\addplot[area legend, line width=0.5pt, draw=black, fill=black, forget plot]
table[row sep=crcr] {%
	x	y\\
	-150	-550\\
	1800	-550\\
}--cycle;

\draw[dashed] (375,-550) -- (375,550);
\draw[dashed] (1125,-550) -- (1125,550);

\addplot[area legend, line width=0.5pt, draw=red, fill=red, forget plot]
table[row sep=crcr] {%
x	y\\
-32.8804653251745	-409.240855018385\\
-30.7591449816148	-407.119534674826\\
32.8804653251745	-470.759144981615\\
30.7591449816148	-472.880465325174\\
-32.8804653251745	-409.240855018385\\
}--cycle;

\addplot[area legend, line width=0.5pt, draw=red, fill=red, forget plot]
table[row sep=crcr] {%
x	y\\
-32.8804653251745	-299.240855018385\\
-30.7591449816148	-297.119534674826\\
32.8804653251745	-360.759144981615\\
30.7591449816148	-362.880465325174\\
-32.8804653251745	-299.240855018385\\
}--cycle;

\addplot[area legend, line width=0.5pt, draw=black, fill=black, forget plot]
table[row sep=crcr] {%
x	y\\
-45	-221.5\\
-45	-218.5\\
45	-218.5\\
45	-221.5\\
-45	-221.5\\
}--cycle;

\addplot[area legend, line width=0.5pt, draw=black, fill=black, forget plot]
table[row sep=crcr] {%
x	y\\
-45	-111.5\\
-45	-108.5\\
45	-108.5\\
45	-111.5\\
-45	-111.5\\
}--cycle;

\addplot[area legend, line width=0.5pt, draw=black, fill=black, forget plot]
table[row sep=crcr] {%
x	y\\
-45	-1.5\\
-45	1.5\\
45	1.5\\
45	-1.5\\
-45	-1.5\\
}--cycle;

\addplot[area legend, line width=0.5pt, draw=black, fill=black, forget plot]
table[row sep=crcr] {%
x	y\\
-45	108.5\\
-45	111.5\\
45	111.5\\
45	108.5\\
-45	108.5\\
}--cycle;

\addplot[area legend, line width=0.5pt, draw=black, fill=black, forget plot]
table[row sep=crcr] {%
x	y\\
-45	218.5\\
-45	221.5\\
45	221.5\\
45	218.5\\
-45	218.5\\
}--cycle;

\addplot[area legend, line width=0.5pt, draw=red, fill=red, forget plot]
table[row sep=crcr] {%
x	y\\
-30.7591449816148	297.119534674826\\
-32.8804653251745	299.240855018385\\
30.7591449816148	362.880465325174\\
32.8804653251745	360.759144981615\\
-30.7591449816148	297.119534674826\\
}--cycle;

\addplot[area legend, line width=0.5pt, draw=red, fill=red, forget plot]
table[row sep=crcr] {%
x	y\\
-30.7591449816148	407.119534674826\\
-32.8804653251745	409.240855018385\\
30.7591449816148	472.880465325174\\
32.8804653251745	470.759144981615\\
-30.7591449816148	407.119534674826\\
}--cycle;
\node[align=center]
at (axis cs:150,-440) {1};
\node[align=center]
at (axis cs:150,-330) {2};
\node[align=center]
at (axis cs:150,-220) {3};
\node[align=center]
at (axis cs:150,-110) {4};
\node[align=center]
at (axis cs:150,0) {5};
\node[align=center]
at (axis cs:150,110) {6};
\node[align=center]
at (axis cs:150,220) {7};
\node[align=center]
at (axis cs:150,330) {8};
\node[align=center]
at (axis cs:150,440) {9};
\node[align=center]
at (axis cs:0,600) {i};

\addplot[area legend, line width=0.5pt, draw=black, fill=black, forget plot]
table[row sep=crcr] {%
x	y\\
705	-441.5\\
705	-438.5\\
795	-438.5\\
795	-441.5\\
705	-441.5\\
}--cycle;

\addplot[area legend, line width=0.5pt, draw=black, fill=black, forget plot]
table[row sep=crcr] {%
x	y\\
705	-331.5\\
705	-328.5\\
795	-328.5\\
795	-331.5\\
705	-331.5\\
}--cycle;

\addplot[area legend, line width=0.5pt, draw=black, fill=black, forget plot]
table[row sep=crcr] {%
x	y\\
705	-221.5\\
705	-218.5\\
795	-218.5\\
795	-221.5\\
705	-221.5\\
}--cycle;

\addplot[area legend, line width=0.5pt, draw=black, fill=black, forget plot]
table[row sep=crcr] {%
x	y\\
705	218.5\\
705	221.5\\
795	221.5\\
795	218.5\\
705	218.5\\
}--cycle;

\addplot[area legend, line width=0.5pt, draw=black, fill=black, forget plot]
table[row sep=crcr] {%
x	y\\
705	328.5\\
705	331.5\\
795	331.5\\
795	328.5\\
705	328.5\\
}--cycle;

\addplot[area legend, line width=0.5pt, draw=black, fill=black, forget plot]
table[row sep=crcr] {%
x	y\\
705	438.5\\
705	441.5\\
795	441.5\\
795	438.5\\
705	438.5\\
}--cycle;
\node[align=center]
at (axis cs:900,-440) {1};
\node[align=center]
at (axis cs:900,-330) {2};
\node[align=center]
at (axis cs:900,-220) {3};
\node[align=center]
at (axis cs:900,220) {7};
\node[align=center]
at (axis cs:900,330) {8};
\node[align=center]
at (axis cs:900,440) {9};
\node[align=center]
at (axis cs:750,600) {ii};

\draw[>=stealth,<->] (750,-210) --node[midway,left](){440mm} (750,210);
\draw[>=stealth,<->] (750,-230) --node[midway,left](){110mm} (750,-320);
\draw[>=stealth,<->] (750,220) --node[midway,left](){110mm} (750,320);

\addplot[area legend, line width=0.5pt, draw=black, fill=black, forget plot]
table[row sep=crcr] {%
x	y\\
1635	-219.8299\\
1634.72262	-219.80074\\
1633.89192	-219.7138\\
1632.5133	-219.57052\\
1630.59504	-219.37324\\
1628.1492	-219.12556\\
1625.19054	-218.83126\\
1621.7376	-218.49484\\
1617.81144	-218.12152\\
1613.43654	-217.71652\\
1608.63954	-217.2856\\
1603.45032	-216.83398\\
1597.90074	-216.36778\\
1592.02482	-215.89276\\
1585.8591	-215.41504\\
1579.44156	-214.94074\\
1572.81144	-214.47616\\
1566.01014	-214.02796\\
1559.07906	-213.60334\\
1552.0614	-213.20914\\
1545	-212.8531\\
1537.9386	-212.5426\\
1530.92094	-212.28502\\
1523.98986	-212.08774\\
1517.18856	-211.95742\\
1510.55844	-211.9\\
1504.1409	-211.92052\\
1497.97518	-212.02258\\
1492.09926	-212.20852\\
1486.54968	-212.47924\\
1481.36046	-212.83384\\
1476.56346	-213.2698\\
1472.18856	-213.78334\\
1468.2624	-214.36978\\
1464.80946	-215.023\\
1461.8508	-215.7367\\
1459.40496	-216.5044\\
1457.4867	-217.3198\\
1456.10808	-218.17696\\
1455.27738	-219.07156\\
1455	-220\\
1455.27738	-220.92844\\
1456.10808	-221.82304\\
1457.4867	-222.6802\\
1459.40496	-223.4956\\
1461.8508	-224.2633\\
1464.80946	-224.977\\
1468.2624	-225.63022\\
1472.18856	-226.21666\\
1476.56346	-226.7302\\
1481.36046	-227.16616\\
1486.54968	-227.52076\\
1492.09926	-227.79148\\
1497.97518	-227.97742\\
1504.1409	-228.07948\\
1510.55844	-228.1\\
1517.18856	-228.04258\\
1523.98986	-227.91226\\
1530.92094	-227.71498\\
1537.9386	-227.4574\\
1545	-227.1469\\
1552.0614	-226.79086\\
1559.07906	-226.39666\\
1566.01014	-225.97204\\
1572.81144	-225.52384\\
1579.44156	-225.05926\\
1585.8591	-224.58496\\
1592.02482	-224.10724\\
1597.90074	-223.63222\\
1603.45032	-223.16602\\
1608.63954	-222.7144\\
1613.43654	-222.28348\\
1617.81144	-221.87848\\
1621.7376	-221.50516\\
1625.19054	-221.16874\\
1628.1492	-220.87444\\
1630.59504	-220.62676\\
1632.5133	-220.42948\\
1633.89192	-220.2862\\
1634.72262	-220.19926\\
1635	-220.1701\\
}--cycle;

\addplot[area legend, line width=0.5pt, draw=black, fill=black, forget plot]
table[row sep=crcr] {%
x	y\\
1635	220.1701\\
1634.72262	220.19926\\
1633.89192	220.2862\\
1632.5133	220.42948\\
1630.59504	220.62676\\
1628.1492	220.87444\\
1625.19054	221.16874\\
1621.7376	221.50516\\
1617.81144	221.87848\\
1613.43654	222.28348\\
1608.63954	222.7144\\
1603.45032	223.16602\\
1597.90074	223.63222\\
1592.02482	224.10724\\
1585.8591	224.58496\\
1579.44156	225.05926\\
1572.81144	225.52384\\
1566.01014	225.97204\\
1559.07906	226.39666\\
1552.0614	226.79086\\
1545	227.1469\\
1537.9386	227.4574\\
1530.92094	227.71498\\
1523.98986	227.91226\\
1517.18856	228.04258\\
1510.55844	228.1\\
1504.1409	228.07948\\
1497.97518	227.97742\\
1492.09926	227.79148\\
1486.54968	227.52076\\
1481.36046	227.16616\\
1476.56346	226.7302\\
1472.18856	226.21666\\
1468.2624	225.63022\\
1464.80946	224.977\\
1461.8508	224.2633\\
1459.40496	223.4956\\
1457.4867	222.6802\\
1456.10808	221.82304\\
1455.27738	220.92844\\
1455	220\\
1455.27738	219.07156\\
1456.10808	218.17696\\
1457.4867	217.3198\\
1459.40496	216.5044\\
1461.8508	215.7367\\
1464.80946	215.023\\
1468.2624	214.36978\\
1472.18856	213.78334\\
1476.56346	213.2698\\
1481.36046	212.83384\\
1486.54968	212.47924\\
1492.09926	212.20852\\
1497.97518	212.02258\\
1504.1409	211.92052\\
1510.55844	211.9\\
1517.18856	211.95742\\
1523.98986	212.08774\\
1530.92094	212.28502\\
1537.9386	212.5426\\
1545	212.8531\\
1552.0614	213.20914\\
1559.07906	213.60334\\
1566.01014	214.02796\\
1572.81144	214.47616\\
1579.44156	214.94074\\
1585.8591	215.41504\\
1592.02482	215.89276\\
1597.90074	216.36778\\
1603.45032	216.83398\\
1608.63954	217.2856\\
1613.43654	217.71652\\
1617.81144	218.12152\\
1621.7376	218.49484\\
1625.19054	218.83126\\
1628.1492	219.12556\\
1630.59504	219.37324\\
1632.5133	219.57052\\
1633.89192	219.7138\\
1634.72262	219.80074\\
1635	219.8299\\
}--cycle;
\node[align=center]
at (axis cs:1500,600) {iii};
\node[align=center]
at (axis cs:1680,-220) {3};
\node[align=center]
at (axis cs:1680,220) {7};
\draw[>=stealth,<->] (1500,-210) --node[midway,left](){440mm} (1500,210);
\draw[>=stealth,<->] (1500,230) --node[midway,left](){280mm} (1500,540);
\draw[>=stealth,<->] (1500,-230) --node[midway,left](){280mm} (1500,-540);
\end{axis}
\end{tikzpicture}%
	\caption{Shafts Setup of the 2D active grid as top view on the shafts. Figure \textbf{i} shows a setup with nine shafts, \textbf{ii} uses six shafts and in \textbf{iii} only two shafts of type NACA0009 are used.}
	\label{fig:Shafts_setup}
\end{figure}

The 2D active grid uses up to nine vertical shafts with a span of $s_{shaft}$=800mm and a chord length $c_{shaft}$ depending on the type used. The gauge in between shaft positions is $g_{shaft}$=110mm. A technical drawing of the grid is shown in Fig. \ref{fig:2DGrid_Technical}. To allow high flexibility in the movement of the shafts, each shaft is moved by its own stepper motor. Motion protocols which contain the position for each time step are given to the stepper motors via NI motion control boards. The time between different positions is 20ms. A whole rotation of a shaft is divided into 51200 micro steps resulting in a theoretical step resolution of 0.007$^\circ$/step. Depending on the used shaft type the stepper motors can be operated at an angular velocity between 400 $^\circ/s$ and 950 $^\circ/s$. The angle $\gamma$ describes the orientation of the shaft with respect to the incoming flow (see Fig. \ref{fig:2DGrid_Technical}). The minimum and maximum possible blockage of the grid depends on the used grid configurations, which are presented below.
\\
In this paper, the resulting flow fields of three grid configurations with different shaft types (see Fig. \ref{fig:ShaftsTypes}) will be investigated. Shaft type \textbf{I} is made of flat plates with a chord length of $c_{\text{FP}}$= 90mm and a thickness of $d_{\text{FP}}$= 5mm. The pivot point of these shafts is at $c/2$ to reduce inertia during the rotation. This type was chosen for ease of manufacture and procurement. Shaft type \textbf{II} is made of 3D printed NACA0018 airfoils with a chord length of $c_{\text{NACA0018}}$= 71mm and a thickness of $d_{\text{NACA0018}}$= 12.8mm. For aerodynamic reasons, the pivot point of those airfoils is at $c/4$. The shape is chosen to optimize aerodynamic properties of the shafts. Especially for large $\gamma$, aerodynamic shapes show a more benign behavior than flat plates. Shaft type \textbf{III} is made of 3D printed NACA0009 airfoils with  a chord length of $c_{\text{NACA0009}} $= 180mm and a thickness of $d_{\text{NACA0009}} $= 16.2mm. Just like the NACA0018 airfoils, these are also rotated around $c/4$. This type of shaft is chosen to study the influence of a larger chord.
\\
Beside shaft geometry, also the number of shafts as well as their arrangement can be varied. Fig. \ref{fig:Shafts_setup} shows three configurations examined in this study. Configuration \textbf{i} shows the default setup with all nine shafts installed. This setup can be used for AoI modulations (see Sec. \ref{chap:AoImodulation}) or by using the outer shafts [1,2,8,9] marked in red to modulate the blockage and therefor the longitudinal velocity (see Sec. \ref{chap:velocitymodulation}).\\
The AoI variations shown in \ref{chap:AoImodulation} and the longitudinal variations shown in \ref{chap:velocitymodulation} can be combined to realize simultaneous modulation of these quantities. This is described in Sec. \ref{chap:bothmodulation}. Due to space limitations, only shaft types \textbf{I} and \textbf{II} are considered for this setup.\\
In \textbf{ii} the inner three shafts [4,5,6] are removed. This allows the generation of a vertical gust without inducing any direct disturbances into the flow field. Since flow quality is the focus of this setup, only shaft type \textbf{II} are used.\\
The third configuration \textbf{iii} uses only two shafts at the position [3,7]. In this configuration NACA0009 airfoils \textbf{III} are used. This configuration is the most used in the literature for generating AoI variations. For this reason, it will be included as a reference.

\section{Results}
\label{chap:Inflow}
In order to characterize the grid, the dimensionless reduced frequency 
\begin{equation}
	k = \frac{\pi \ c \ f}{u_\infty}
\end{equation}
is used in all presented results \cite{leishman2006principles}. $c$ denotes the chord length of the considered shaft, $u_\infty$ the inflow velocity and $f$ the frequency of the shaft motion or comparably the frequency of the modulated inflow. Since $k$ considers the frequency of the gust and the free flow velocity, this quantity is ideal for comparing the results of different shaft types and velocities.\\ 

\subsection{Two-dimensionality of the generated flow}
\label{chap:2DInflow}
\begin{figure}[h]
	\centering
	\includegraphics[width=0.99\linewidth]{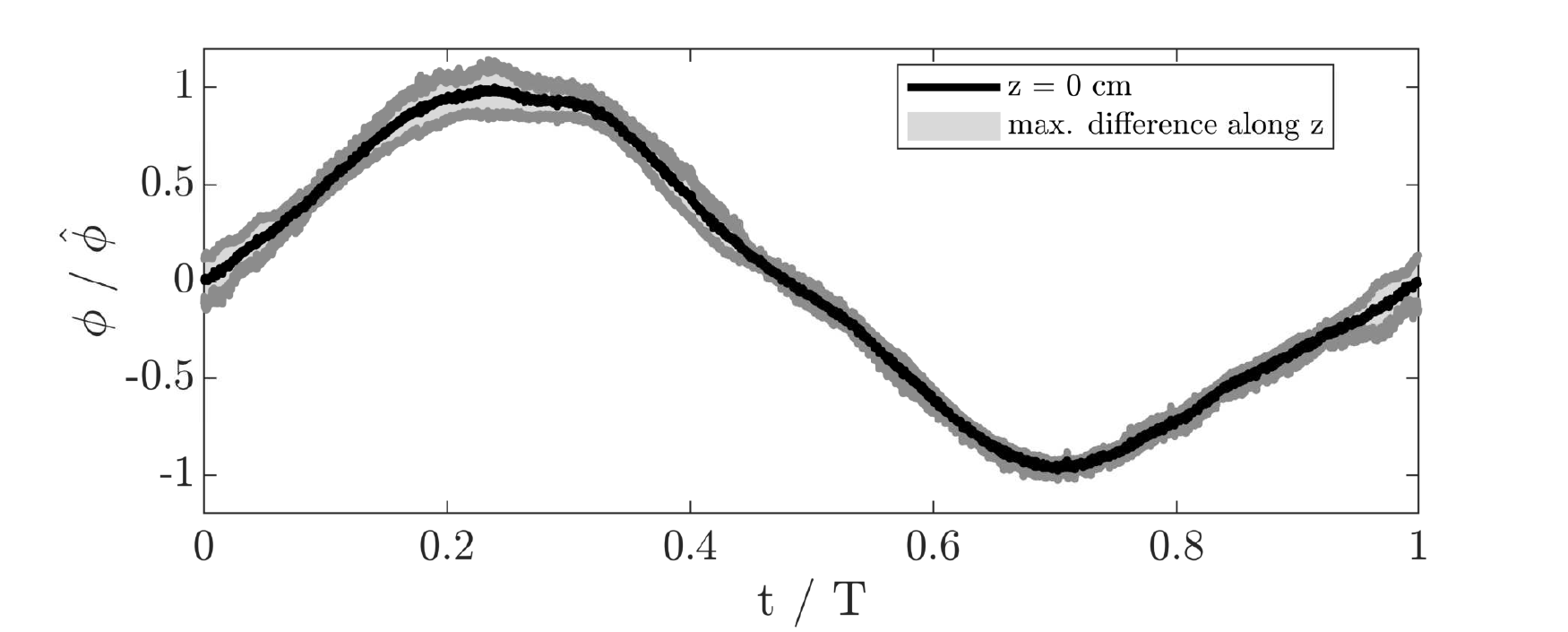}
	\caption[Traversed hot-wire]{Measurement of $\phi$ along the z-direction. The black line represents the $z=0mm$ measurement. The maximum deviation between centerline and measurements at $z=(-50,-100,-150,-200,-250,-300)$mm is given by the error bar in gray. The measurements are performed with setup \textbf{II/i} at $k=0.08$ and $\hat{\gamma} = 7.5^\circ$ amplitude. The results are normalized by the AoI amplitude $\hat{\phi}$.}
	\label{fig:2dcheck}
\end{figure}

\begin{figure}[h]
	\centering
	\includegraphics[width=0.99\linewidth]{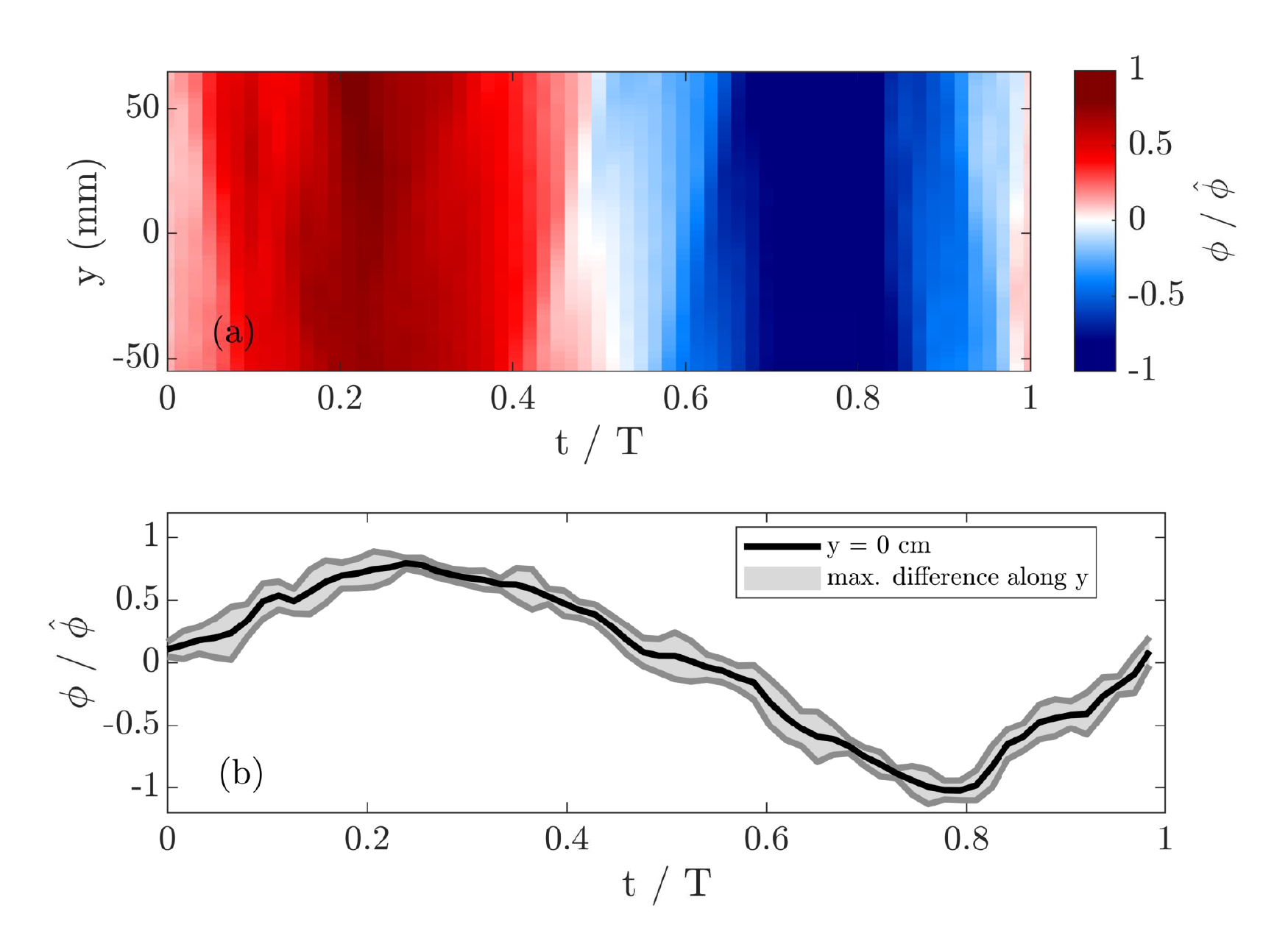}
	\caption[Inflow from PIV]{(a) Phase average PIV results of the angle of incidence modulation in y-direction. (b) Extracted time series from (a). The black line represents $y=0$mm. The maximum deviation between centerline and the time series from $y=-55$mm to $y=65$mm (80 time series) is given by the error bar in gray. The measurements are performed with setup \textbf{II/i} at $k=0.13$ and $\hat{\gamma} = 7.5^\circ$ amplitude. The results are normalized by the AoI amplitude $\hat{\phi}$.}
	\label{fig:PIVInflow}
\end{figure}

Main motivation for building the 2D active grid is to get rid of the three-dimensional flow along the z-direction induced by the previously used 3D active grid \cite{traphan2020dynamic}. Therefore, the focus of this part is to verify the homogeneity of incident gust along the wind tunnel height. This is checked by two simultaneous X-wire measurements. One X-wire is positioned on the center line of the test section, while a second X-wire is moved down in z-direction by steps of 5cm, symmetry in span-wise direction assumed. For these comparison measurements the setup \textbf{II/i} is chosen and operated with a reduced frequency of $k = 0.08$ with an amplitude of $\hat{\gamma} = 7.5^\circ$. The results of the phase averaged measurement are shown in Fig. \ref{fig:2dcheck}. The results from the centerline (black line) shows very small deviation from the other measurements in the z-direction shown as gray error bars.
\\
To check the homogeneity in horizontal direction in the later measurement volume, PIV measurements are performed. The method was chosen to get a higher spatial resolution of the interaction zone of neighboring shafts. Fig. \ref{fig:PIVInflow}(a) shows the phase averaged resulting AoI along the y-direction. The figure shows a homogeneous AoI modulation in horizontal direction of the test section. In addition, Fig. \ref{fig:PIVInflow}(b) shows a comparison of the along y-direction extracted time series with respect to the centerline measurement $y=0$mm.
\\
Both data show a non-changing flow along the vertical and horizontal direction. The shape and the standard deviation of the AoI curves along z- and y-direction is comparable. The results also do not show a phase shift between the different measuring points. This is a clear indication of a purely two-dimensional flow generated by the 2D active grid setup.

\subsection{Angle of incidence modulation}
\label{chap:AoImodulation}
\begin{figure}[h]
	\centering
	\includegraphics[width=0.99\linewidth]{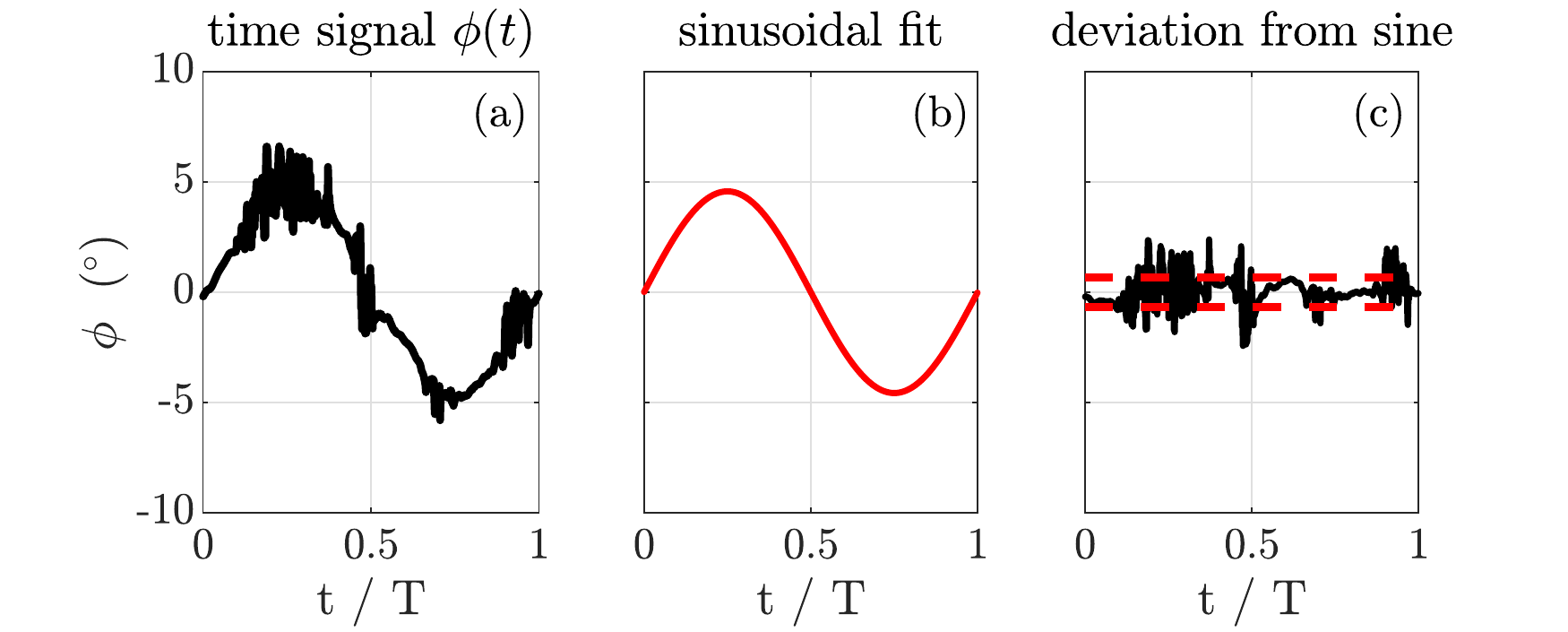}
	\caption[Error estimation]{(a) Raw time signal of the AoI fluctuation $\phi(t)$. (b) Sinusoidal fit of the data shown in (a). (c) Difference of (a) and (b) with the resulting standard deviation plotted as red dashed lines.}
	\label{fig:ErrorEstimation}
\end{figure}

\begin{figure*}[h]
	\centering
	\includegraphics[width=0.99\linewidth]{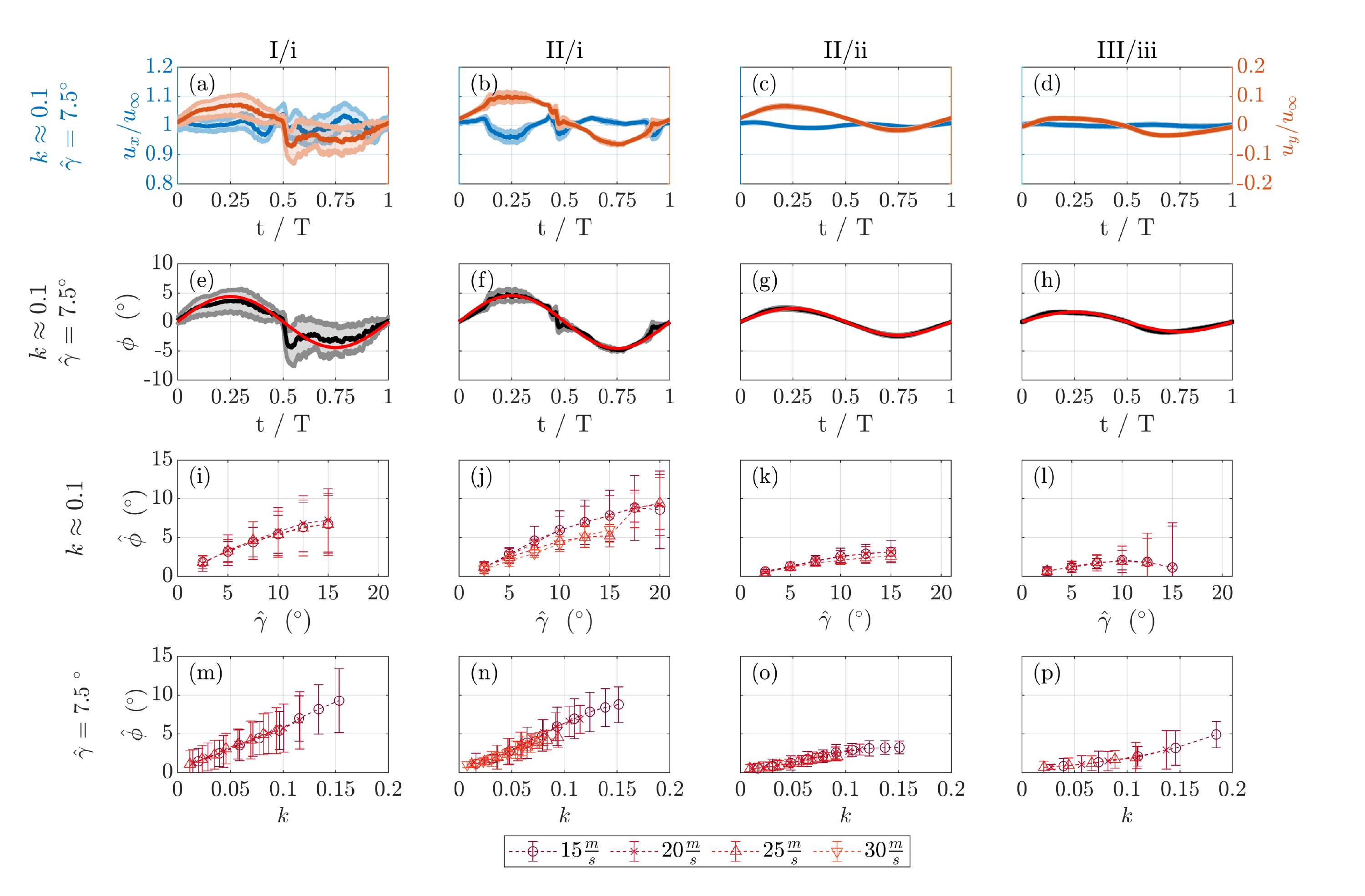}
	\caption[Comparison of AoI variations for different setups]{Comparison of the flow generated by different setups. Each column represents one setup. First line shows the phase averaged time signal of measured velocity components for a grid amplitude of $\hat{\gamma} = 7.5^\circ$ at a reduced frequency of $k\approx 0.1$ ((a)-(d)). The error bars represent the standard deviation of the phase average. Second line shows the resulting phase averaged AoI ((e)-(h)). The error bars represent the standard deviation of the phase average. Third line shows the resulting AoI fluctuations $\hat{\phi}$ for different grid amplitudes $\hat{\gamma}$ at a fixed reduced frequency of $k\approx 0.1$ ((i)-(l)). In the fourth line the resulting AoI fluctuation for different reduced frequencies of the shafts at fixed amplitude of $\hat{\gamma}=7.5^\circ$ is plotted ((m)-(p)).}
	\label{fig:AoIAll}
\end{figure*}

\begin{figure}[h]
	\centering
	\includegraphics[width=0.75\linewidth]{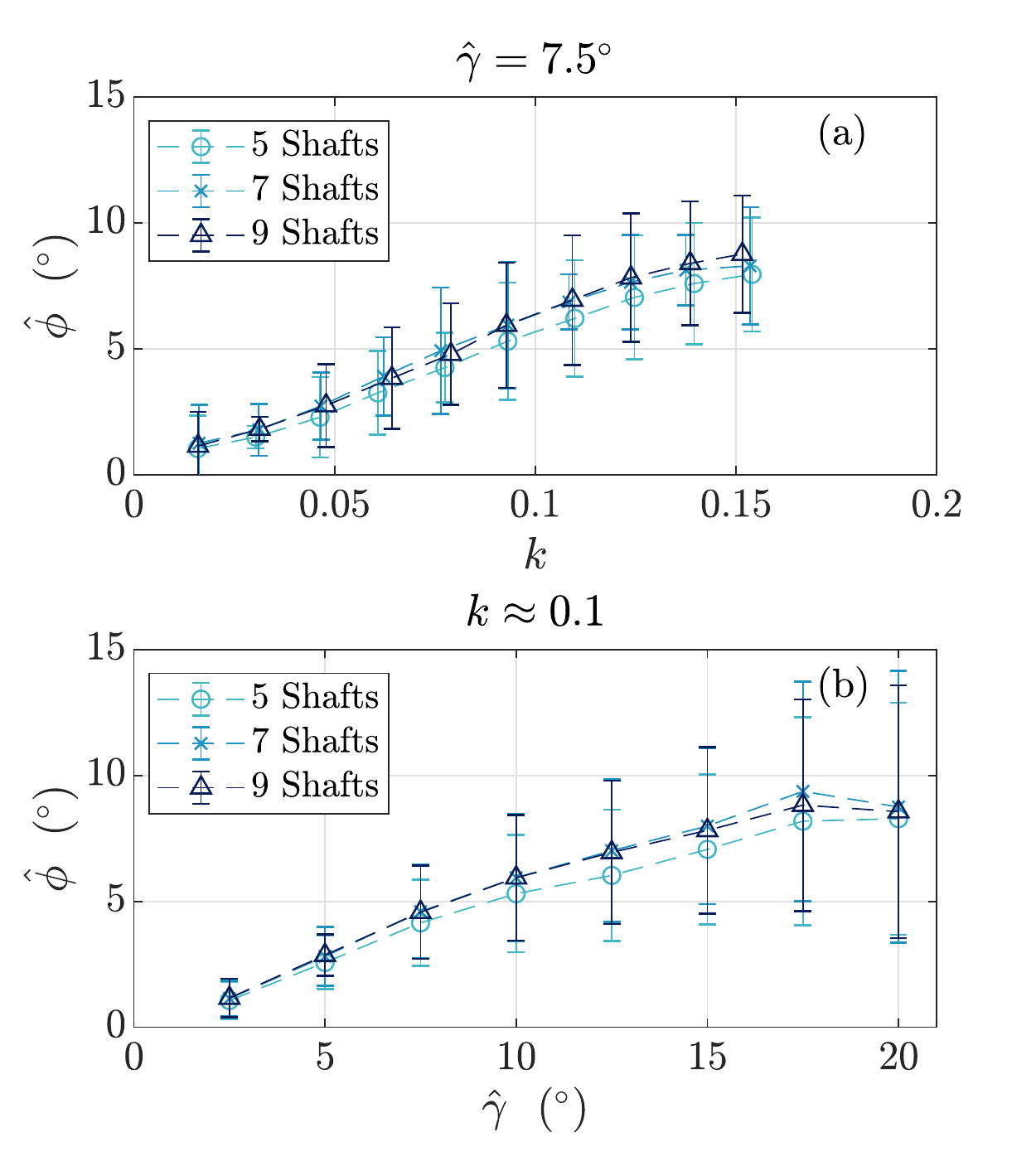}
	\caption[Quantaty of shafts]{Variation of the number of used shafts in setup \textbf{II/i} at fixed wind speed of 15m/s. (a) Resulting AoI magnitude $\hat{\phi}$ for different reduced frequencies $k$ for five, seven or nine shafts. (b) Resulting AoI magnitudes for different shaft amplitudes $\hat{\gamma}$ at fixed reduced frequency of $k\approx 0.1$.}
	\label{fig:ShaftQuantaty}
\end{figure}

This section will discuss different setups with respect to the generated flow qualities. To cover a broad operation range of the setups the parameters given in table \ref{tab:Parameter} are tested. The data are generated using sinusoidal movements of the grid shafts with different frequencies $f$ and amplitudes $\hat{\gamma}$ of the shafts. 
\begin{table}[htbp]
	\begin{center}
		\caption{Parameter used during the characterization of the 2D active grid for different setups.}
		\label{tab:Parameter}       
		\begin{tabular}{lcccc}
			\hline\noalign{\smallskip}
			Setup & \multicolumn{1}{l}{f (Hz)} & \multicolumn{1}{l}{$\Delta$ f (Hz)} & \multicolumn{1}{l}{$\hat{\gamma}$ ($^\circ$)} & \multicolumn{1}{l}{$\Delta \hat{\gamma} $ ($^\circ$)} \\
			\midrule
			\textbf{I/i} & 1-8 & 1 & 2.5-15 & 2.5 \\
			\textbf{II/i} & 1-10 & 1 & 2.5-20 & 2.5 \\
			\textbf{II/ii} & 1-10 & 1 & 2.5-15 & 2.5 \\
			\textbf{III/iii} & 1-5 & 1 & 2.5-15 & 2.5 \\
			\noalign{\smallskip}\hline
		\end{tabular}
	\end{center}
\end{table}\\
Fig. \ref{fig:AoIAll} shows an overview of chosen flow variations out of the measured parameter space to compare the setups. The lines show the following: 
\begin{enumerate}
	\item Resulting phase averaged velocity components $u_x$ and $u_y$
	\item Resulting phase averaged angle of incidence (AoI) variation $\phi$
	\item Resulting AoI magnitude $\hat{\phi}$ as a function of the shaft amplitude $\hat{\gamma}$
	\item Resulting AoI magnitude $\hat{\phi}$ as a function of the reduced frequency of the grid $k$
\end{enumerate}
The fourth and fifth lines reflect only a cut through the tested parameter space. To obtain the error bars shown in Fig. \ref{fig:AoIAll}(i-p), a sinusoidal fit is subtracted from the measured AoI time series. From the remaining fluctuations, the standard deviation is calculated. This gives a measure of the deviation from the perfect sinusoidal signal. This procedure is exemplified in Fig. \ref{fig:ErrorEstimation}.
\\
All four tested setups show differences between each other. In general, it is possible to distinguish between two categories. Configuration \textbf{i} with axes on the centerline and configurations \textbf{ii} and \textbf{iii}, where the centerline is vacant. Configuration \textbf{i} produces larger fluctuations in the $u_x$ component. The fluctuations reach peak values up to 5$\%$ of the inflow velocity. For configuration \textbf{ii} and \textbf{iii} they are smaller than 1$\%$ (see Fig. \ref{fig:AoIAll}(a-d)). In return, \textbf{ii} and \textbf{iii} do not reach very large $u_y$ fluctuations. Furthermore, both setups of \textbf{i} show a structure at $t/T = 0.5$, which is not present in the other ones. This effect can be traced back to the wake of the centerline shaft, which is only present in configuration \textbf{i}, and passes the measurement point every time when  the shaft angle $\gamma$ is 0$^\circ$. All these effects can also be seen in the resulting AoI variations $\phi$ in Fig. \ref{fig:AoIAll}(e-h). Due to the higher $u_y$ variation configuration \textbf{i} generates larger AoI variations. The resulting AoI variation reflects for all cases the trend of the $u_y$ component and is only slightly influenced by the $u_x$ component. Fig. \ref{fig:AoIAll}(e-h) also contains a sinusoidal fit for the measured AoI. This fit matches the measured time series very well in Fig. \ref{fig:AoIAll}(f-h). Fig. \ref{fig:AoIAll}(e) shows a deviation from the pure sinusoidal form. Another difference between the tested configurations are the standard deviations of the phase averages, which are plotted as error tubes. While these are very small for \textbf{II/i}, \textbf{II/ii} and \textbf{III/iii}, which indicates a very clean flow modulation without turbulence generation, setup \textbf{I/i} shows significantly larger standard deviations.\\
The last two lines of Fig. \ref{fig:AoIAll} represent the resulting AoI magnitudes $\hat{\phi}$ for a fixed reduced frequency of the grid shafts $k \approx 0.1$ (see Fig. \ref{fig:AoIAll}(i-l)) or for a fixed grid amplitude $\hat{\gamma} = 7.5^\circ$ (see Fig. \ref{fig:AoIAll}(m-p)). The magnitude of the AoI modulation is hereby extracted from the sinusoidal fit. \\
The comparison of the resulting amplitudes shows larger AoI variations for configuration \textbf{i}, as already noted before. The amplitudes are larger by a factor of three compared to configuration \textbf{ii} and \textbf{iii}. However, the error bars for \textbf{i} are larger, indicating greater deviations from the sinusoidal shape. For configuration \textbf{i} the shafts of type \textbf{II} show a significant reduction of the error bar size. By comparing \textbf{ii} and \textbf{iii}, \textbf{iii} shows larger error bars. In general, the error bars increase with rising shaft amplitude or with higher reduced frequency for all setups. The amplitudes show the same dependence on the reduced frequency for different wind speeds. In addition, for $k \approx 0.15$, amplitudes of up to $\hat{\phi} = 15^\circ$ can be achieved using the setup \textbf{II/ii}. Figures showing the entire parameter space can be found in the supplementary file.\\
In summary, two important findings can be mentioned here. On the one hand, configuration \textbf{i} generates not only $u_y$ fluctuations but also fluctuation in $u_x$, which are less pronounced in setup \textbf{ii} and \textbf{iii}. However, these fluctuations can be reduced by using shaft type \textbf{II}. On the other hand, configuration \textbf{ii} and \textbf{iii} deliver comparable results, but the higher number of shafts in \textbf{ii} gives better flow qualities and higher amplitudes. The measurements result in a look up table for the AoI amplitude $\hat{phi}$ in dependence of the shaft amplitude $\hat{\gamma}$ and the reduced frequency $k$. By this the required shaft amplitude $\hat{\gamma}$ can be determined for later experiments to match desired AoI amplitudes $\hat{\phi}$ at a desired reduced frequency $k$.
\\
In addition to the previous investigation, the influence of not using all shafts in setup \textbf{II/i} to redirect the flow is further investigated. During characterization, the outer shafts are fixed at $\gamma = 0^\circ$ position. Fig. \ref{fig:ShaftQuantaty} shows the resulting magnitudes $\hat{\phi}$ for a wind speed of 15m/s. Fig. \ref{fig:ShaftQuantaty}(a) represents the magnitude relative to the reduced frequency and Fig. \ref{fig:ShaftQuantaty}(b) the magnitude relative to the shaft amplitude. The results for nine shafts are the same as shown in Fig. \ref{fig:AoIAll}(j) and \ref{fig:AoIAll}(n). Seven shafts mean shafts [1,9] are not moving, for five shafts [1,2,8,9] are not moving. The curves do not indicate a large influence of the outer shafts on the generated flow. The amplitude decreases only slightly when using fewer shafts. The shape of the curves also remains completely unchanged.

\subsection{Velocity modulation}
\label{chap:velocitymodulation}
\begin{figure}[htbp]
	\centering
	\includegraphics[width=0.75\linewidth]{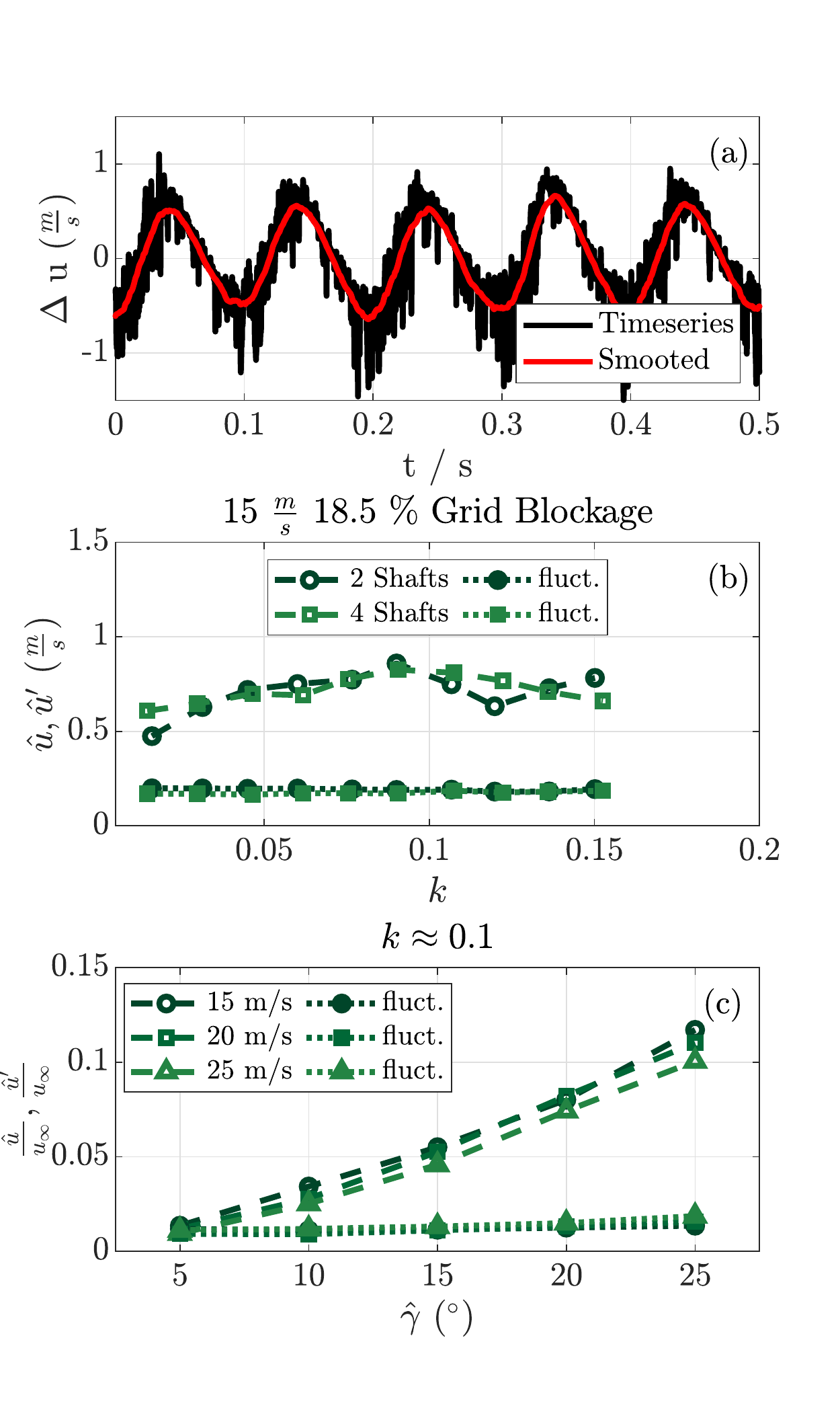}
	\caption[Velocity fluctuation time series]{Overview over the longitudinal gusts generated by setup \textbf{II/i}. (a) Time series of velocity fluctuations induced by increasing the blockage using outer shafts. The reduced frequency is $k = 0.11 $ at an inflow velocity of $20$m/s. The red line shows the moving average over 500 samples of the time series. (b) Velocity variation amplitudes $\hat{u}$ for a fixed blockage ratio of 18.5$\%$. The filled symbols represent the standard deviation $\hat{u}'$ around the modulated flow. (c) Normalized velocity variation amplitudes for reduced frequency of $k \approx 0.1$ at different free stream velocities using four outer shafts. The filled symbols represent the standard deviation around the modulated flow.}
	\label{fig:TS_Vel_Fluc}
\end{figure}
The following chapter investigates how strong longitudinal gusts can be imposed. Using the 2D active grid the velocity modulations up to $10$Hz can be imposed upstream of the airfoil by using the outer shafts [1,2,8,9]. Setup \textbf{II/i} is chosen for the following experiments.\\
During investigation, the inner shafts remain unchanged at the $\gamma=0^\circ$ position, while the outer shafts perform a sinusoidal movement. The motion starts at half of the desired amplitude with the shafts trailing edge pointing towards the wind tunnel walls and is then moved between $\gamma=0^\circ$ and two times the amplitude $\hat{\gamma}$. An exemplary time series resulting from such a motion is given in Fig. \ref{fig:TS_Vel_Fluc}(a) where four outer shafts are used to generate such a gust. The amplitude of the movement is $\hat{\gamma}= 10^\circ$ resulting in a velocity modulation of $\hat{u} = 0.6$m/s at an inflow velocity of $20$m/s. The shafts move with a frequency of $f_{Grid} = 10\rm{Hz}$ corresponding to a reduced frequency of $k = 0.11 $. The time series shows a periodic behavior, but the shape of the raw signal is not perfect sinusoidal. This problem arises due to the inertia of the air flow itself. By increasing the blockage, a speed up is achieved. However, when the grid reopens the flow does not recover immediately. This is the reason for the increased fluctuations in the valleys of the signal.\\
In the Fig. \ref{fig:TS_Vel_Fluc}(b) a comparison of the achievable velocity modulation for a blockage of $18.5\%$ is shown. Two cases with a different number of shafts in motion ([1,9] and [1,2,7,9]) are used. The hollow points in the plot represent the amplitude of the velocity modulation $\hat{u}$ from a sinusoidal fit. The filled points represent the fluctuations $\hat{u}'$ around the velocity modulation. The fluctuations are calculated by subtracting the smoothed curve (see Fig. \ref{fig:TS_Vel_Fluc}(a) red line) from the time series and calculating the standard deviation of the remaining signal (comparable to Fig. \ref{fig:ErrorEstimation}). The amplitude of the velocity modulation $\hat{u}$ and the fluctuations $\hat{u}'$ seem to be independent from the reduced grid frequency $k$, which is given as horizontal axis. Furthermore, this plot emphasizes a dependency of the amplitude $\hat{u}$ just on the blockage and not on the number of shafts which are used for generating the blockage. \\
Fig. \ref{fig:TS_Vel_Fluc}(c) shows the induced velocity modulation for three different inflow velocities over the amplitude of the shafts $\hat{\gamma}$. The resulting velocity modulations are normalized by the inflow velocity $u_\infty$, to compare the plots. The filled symbols represent again the fluctuations $\hat{u}'$ around the modulation. Similar behaviors are found for all wind speeds. Hence, the velocity modulation is independent of the inflow velocity and only depending on the induced blockage. This is represented by the monotonic increase of the velocity modulation when the amplitude of the shafts is also increased. Furthermore, the fluctuations $\hat{u}'$ are constant over the entire investigated range.\\
In summary Fig. \ref{fig:TS_Vel_Fluc}(b) and \ref{fig:TS_Vel_Fluc}(c) show a broad range of possible velocity fluctuations and only a dependency on the blockage of the grid. Neither the frequency of the grid nor the inflow velocity has an impact on the resulting fluctuations. Also, the number of shafts used is not critical for the generation of velocity fluctuations. A plot of the other tested parameters can be found in the supplementary file.

\subsection{Combination of AoI and velocity modulation}
\label{chap:bothmodulation}
\begin{figure}[htb]
	\centering
	\includegraphics[width=0.75\linewidth]{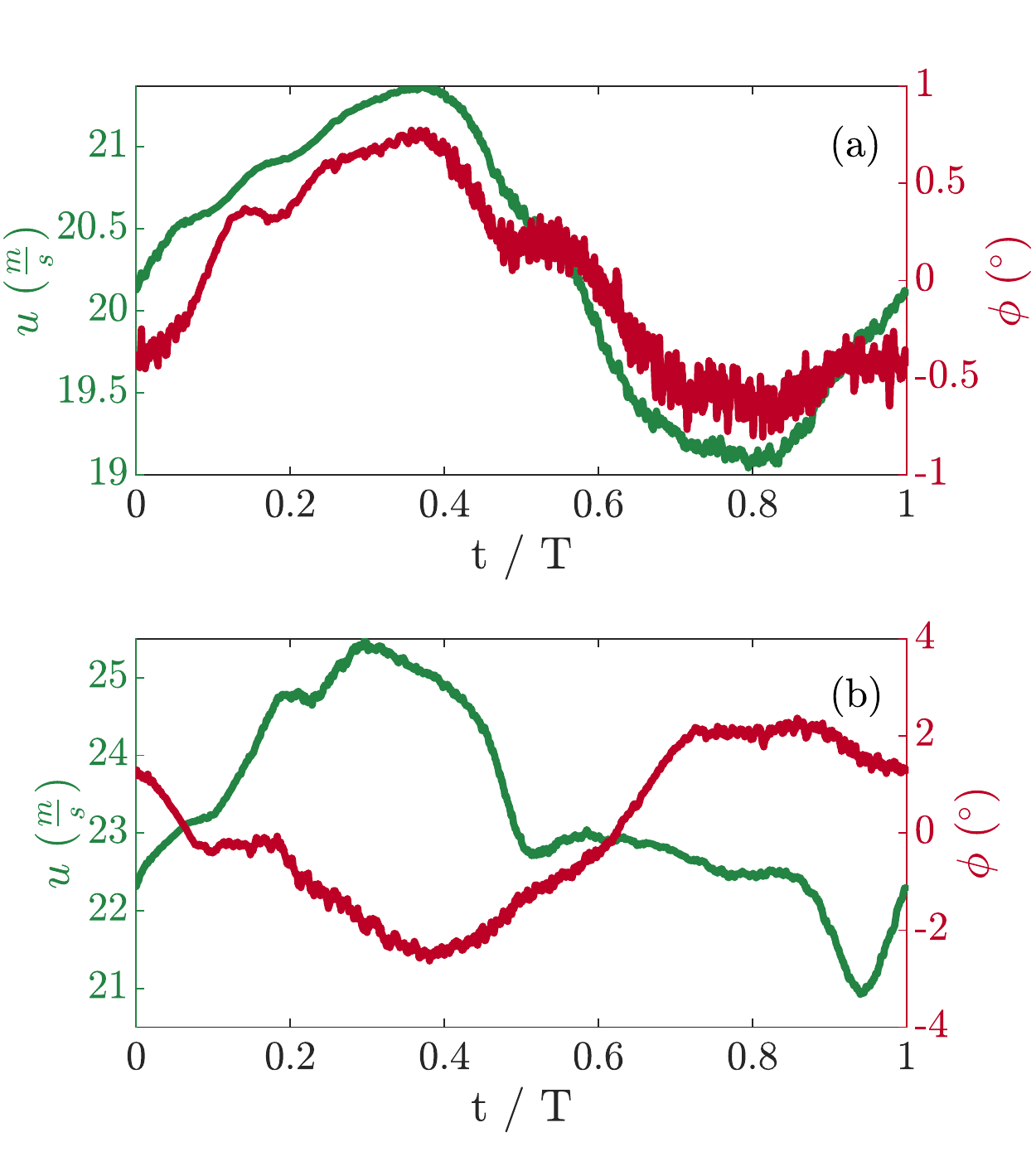}
	\caption[In phase variations]{(a) Example of simultaneous in phase variations of inflow velocity $u$ and angle of incidence $\phi$. (b) Example of simultaneous out of phase variations of inflow velocity $u$ and angle of incidence $\phi$.}
	\label{fig:InPhase}
\end{figure}

Next it will be shown how angle of incidence and velocity modulations can be combined using the 2D active grid with setup \textbf{I/i}. Fig \ref{fig:InPhase}(a) shows the variation of the inflow velocity $u$ in green on the left ordinate and the AoI fluctuations in red on the right ordinate. The variations are a superposition of the fluctuations presented in the previous chapters. The fluctuations shown are in phase, i.e., the angle of incidence is maximum when the velocity reaches its maximum. \\
The opposite phase can be seen in Fig. \ref{fig:InPhase}(b), where the angle of incidence is half a period ahead of the velocity fluctuation. This emphasizes the flexibility of this setup. The phase shift between the angle of incidence and velocity fluctuation can be varied completely arbitrarily. The deviations from a perfect sine can be explained by the chosen setup \textbf{I/i}.
\\
In summary, it was shown in chapter \ref{chap:Inflow} how the generated AoI amplitude depends on the reduced frequency $k$ of the shafts. Also, the amplitude can be changed by the shaft amplitude $\hat{\gamma}$. The dependence of $k$ is not related to the velocity. This enables the extraction of a look up table from given characterization. It could also be shown that longitudinal fluctuations can be generated with the grid, which are hardly dependent on the frequency, but only on the blockage. Last, it was shown that the vertical and longitudinal gusts can also be combined arbitrarily.

\section{Application of 2D grid for airfoil measurements}
\label{chap:PIV}
\begin{figure}[h]
	\centering
	\includegraphics[width=0.75\linewidth]{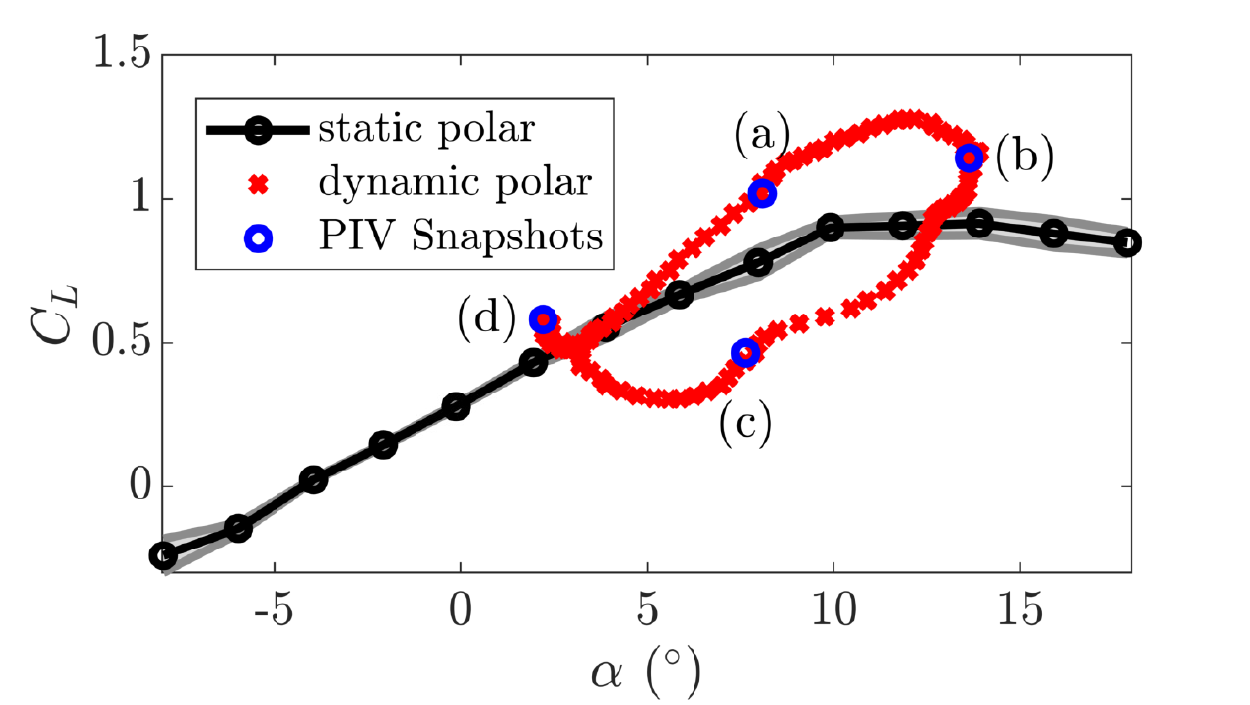}
	\caption[Dynamic polar of an airfoil]{Measured forces of an airfoil for the static case (black) and the dynamic case (red) over the angle of attack $\alpha$. The blue circles indicate the positions of the PIV snapshots shown below on the dynamic stall cylce.}
	\label{fig:DynamicPolar}
\end{figure}

The characterized flows are intended to be used for studies on the dynamic stall of airfoils. Therefore, an application will be shown here to verify that the presented approach can induce a dynamic stall on an airfoil. A Clark Y-profile with a chord length of $c_{Airfoil} = 180$mm is used for this investigation. The experimental setup shown in \ref{chap:Windtunnel} with the active grid is used to induce the dynamic inflow and measure the resulting forces. Other measurements using this setup were presented in \cite{Wei_2019}, \cite{wester2018high}and \cite{Wester2018POD}.\\ 
For this experiment, a pure AoI variation is produced by the setup \textbf{I/i} generating an inflow amplitude of $\hat{\phi} = \pm 6^\circ$. The airfoil is turned to a static geometric angle of attack of $\alpha = 8^\circ$ which leads together with the inflow to a dynamic inflow cycle of $\alpha_{Inflow} \in [2,14]^\circ$.\\ 
Since the airfoil is not turned, no inertial forces occur and therefore no dynamic corrections are necessary. Fig. \ref{fig:DynamicPolar} shows the static polar of the airfoil (measured beforehand under laminar conditions) and the dynamic lift induced by the dynamic inflow angle variation. The sinusoidal inflow has a frequency of $f_{Inflow} = 8 \rm{Hz}$ resulting in a reduced frequency of $k_{Airfoil} = 0.3$, which is a highly unsteady inflow condition. This is also represented by the typical dynamic stall loop, which shows the typical lift overshoot for increasing $AoA$, a sudden drop of lift when the angle does not increase further and the undershoot of lift during the decreasing angle of attack until the minimum is reached and the cycle starts again.
\\
The advantage of this experimental approach presented here is the quite easy way to perform temporal and spatial resolved stereoscopic PIV measurements (see also \cite{wester2018high}) in combination with force measurements. This is only possible, because the airfoil is static and therefore the light sheet and cameras do not need to be adjusted for this large angular range. This setup also enables a measurement which is not phase locked, but fully time resolved.
\begin{figure}[h]
	\centering
	\includegraphics[width=0.85\linewidth]{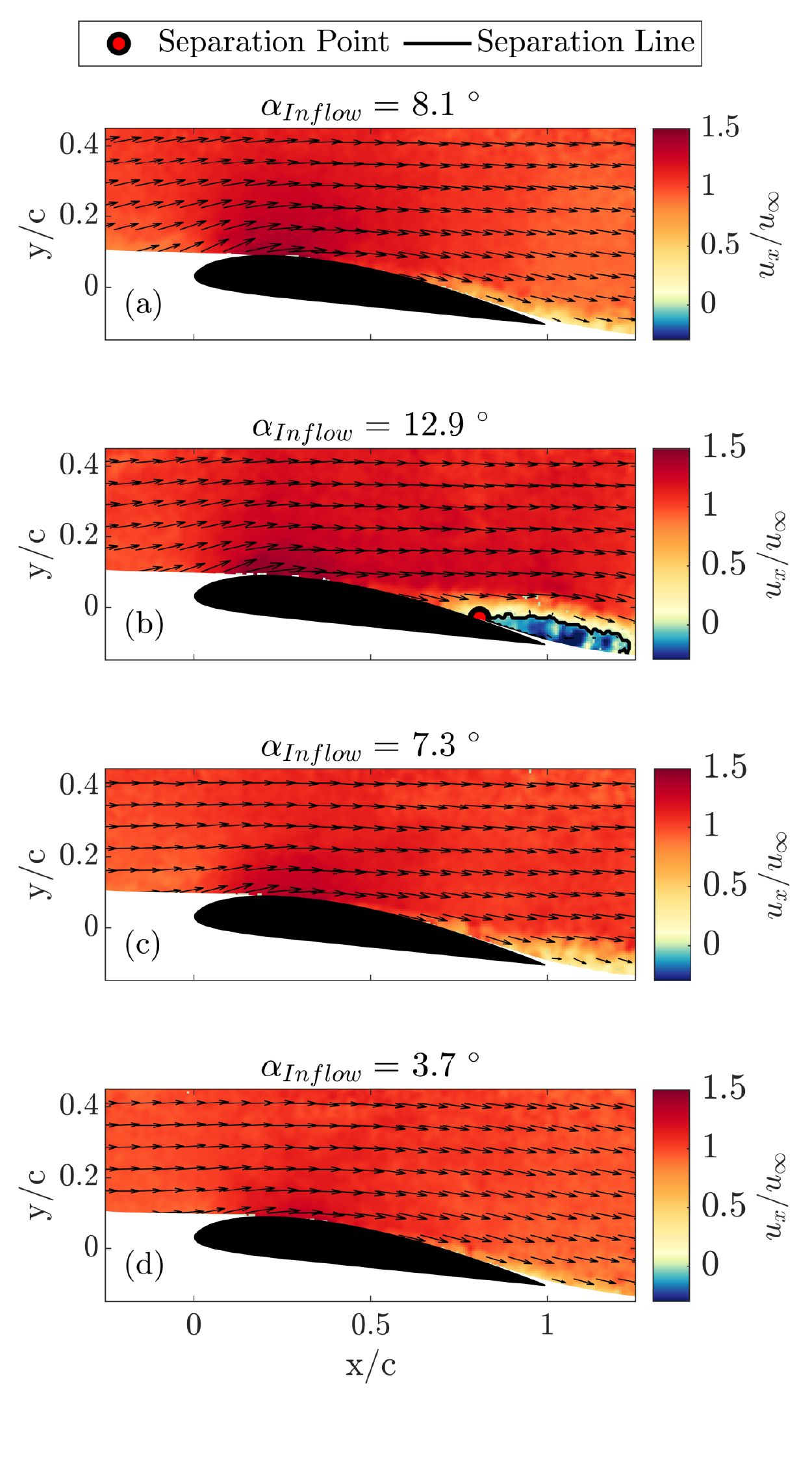}
	\caption[PIV Snapshot]{PIV Snapshots of the flow around an airfoil under dynamic inflow conditions. The pictures correspond to the situations marked in Fig. \ref{fig:DynamicPolar}}
	\label{fig:PIVSnapshot}
\end{figure}\\
Four chosen snapshots of the PIV measurement are shown in Fig. \ref{fig:PIVSnapshot}. The flow field is taken from the same measurement, which resulted in the dynamic stall cycle shown in Fig. \ref{fig:DynamicPolar}. The snapshots show the flow around the airfoil at different angle of attack. A typical behavior of the flow can be observed since the flow stays attached during the rising angle of attack in Fig. \ref{fig:PIVSnapshot}(a), shows a flow separation at the trailing edge at the maximum angle in Fig. \ref{fig:PIVSnapshot}(b), reattaches when the angle of attack decreases again in Fig. \ref{fig:PIVSnapshot}(c) and is also attached at the minimum angle in Fig. \ref{fig:PIVSnapshot}(d). For the separation in Fig. \ref{fig:PIVSnapshot}(b) the area is limited to the trailing edge, the point where the flow separates can also be located and is plotted as a red dot into the picture.\\
This chapter illustrates that the desired flow can be generated with the 2D active grid and the dynamic stall effect can be investigated on an airfoil. The force measurements and the PIV measurements show the typical effects already known from other studies. In contrast to such investigations, it is much easier to perform these experiments with the stationary airfoil.

\section{Conclusion}
\label{chap:Conclusion}
In this study an experimental approach to modulate the inflow in a highly dynamic manner is presented. The 2D active grid enables the generation of purely two-dimensional flows with desired gust amplitude and frequency in vertical direction. With the use of different shaft types, the quality of produced flow can be improved. The 2D active grid shows significantly larger amplitude ranges than previous studies in literature. This is shown as red and orange areas in the Fig. \ref{fig:Lit_Comp}. Red presents large AoI amplitudes generated by configuration \textbf{i} and orange the area with very clean AoI variations generated by configurations \textbf{ii} and \textbf{iii}. This shows the possibility to use the grid for experiments with very high demands on the flow quality as well as for experiments where the flow quality does not have to be perfect and a large AoI amplitude is in focus. As an additional feature, the possibility to generate longitudinal gusts with the grid is presented. Thereby, a dependence on the generated blockage and relative variations independent of the wind speed is shown. This applies to both the shaft frequency and the shaft amplitude. Furthermore, a combination of AoI and velocity variations with arbitrary phase shift could be shown.\\
When designing a 2D active grid the reduced frequency of the grid shafts during experiments needs to be considered. The presented results emphasize the importance of this parameter for the resulting flow field. Furthermore, the following rules can be derived for the design of a 2D active grid: Aerodynamic profiles should be used for the generation of clean AoI variations. For large AoI variations, shafts must be used directly in front of the object under investigation, since passive influence via the sides does not have a large effect. However, it is sufficient to provide half of the wind tunnel width with shafts without strongly reducing the amplitude. For higher requirements on flow quality with lower background turbulence, shafts on the sides of the inlet are recommended. Several small shafts show a better result than a few shafts with larger chord lengths.\\
The shown possibilities to produce various inflow conditions are interesting to generate different flow situations in a wind tunnel and to investigate them under reproducible conditions. This includes for instance the typical flow situations like dynamic stall on a rotor blade of a wind turbine caused by e.g. gusts, yaw misalignment or tower shadow. It is important to understand these effects in detail to incorporate this knowledge into the design of new wind turbines, so the lifetime of such can be extended despite the extreme operating conditions often occurring. The 2D active grid can also be used for other research areas where rapid inflow changes occur. Examples are helicopter aerodynamics or micro air vehicles (MAV) as well as experiments on flying creatures, on whose wings dynamic effects typically play an important role. This emphasizes the broad applicability of the device presented here.

\section{Acknowledgments}
The present investigations were performed within the "Wind Turbine Load Control under Realistic Turbulent In-Flow Conditions" (PE 478/15-2 \& HO 50272-2) project. The authors
gratefully acknowledge the German Research Foundation (DFG) for funding the studies.\\
The authors also want to thank Lars Kröger and Piyush Singh for their support during the measurement campaigns.\\
The authors would also like to thank Cameron Tropea for fruitful discussions during the design phase of the 2D active grid.

\bibliographystyle{spphys}
\bibliography{references}

\end{document}


\title{How to design a 2D active grid for dynamic inflow modulation
}


\author{Tom T. B. Wester \and 
	Johannes Krauss \and 
	 Lars Neuhaus \and 
	  Agnieszka Hölling \and 
	   Gerd Gülker \and 
	    Michael Hölling \and 
	     Joachim Peinke   
}


\institute{T.T.B. Wester \at
	ForWind, Institute of Physics, University of Oldenburg\\
              Küpkersweg 70 \\
              Tel.: +49-441-7985023\\
              Fax: +49-441-7985099\\
              \email{tom.wester@uni-oldenburg.de}           
}

\date{Received: date / Accepted: date}
\section*{Supplementary Material}
This supplementary material shows the measured parameter range for the angles of attack $\hat{\gamma}$, as well as the reduced frequencies $k$ of the shafts.\\
In the first section, the results for angle of incidence (AoI) amplitudes are presented. In the second section the possible velocity variations are shown. 
\section*{Angle of Incidence Modulation}
\begin{figure*}[h]
	\centering
	\includegraphics[width=0.99\linewidth]{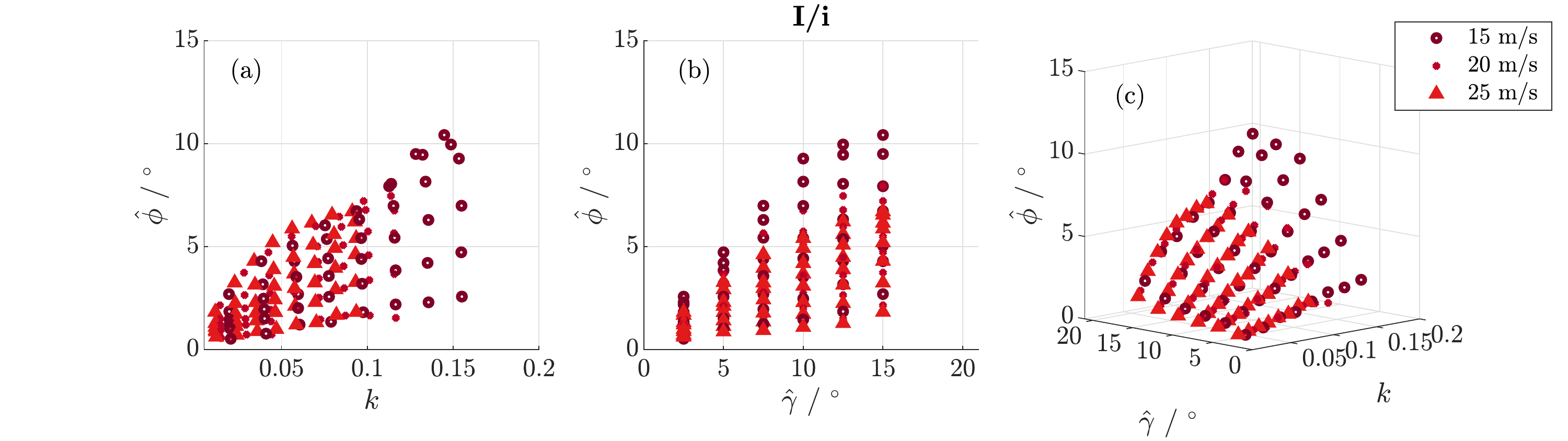}
	\caption[Flat plates all parameters]{Resulting AoI amplitudes for all measurements done with setup \textbf{I/i}. (a) Depending on the reduced frequency $k$. (b) Depending on the shaft amplitude $\hat{\gamma}$. (c) 3D view of results.}
	\label{fig:Flat_Plates_all}
\end{figure*}

\begin{figure*}[h]
\centering
\includegraphics[width=0.99\linewidth]{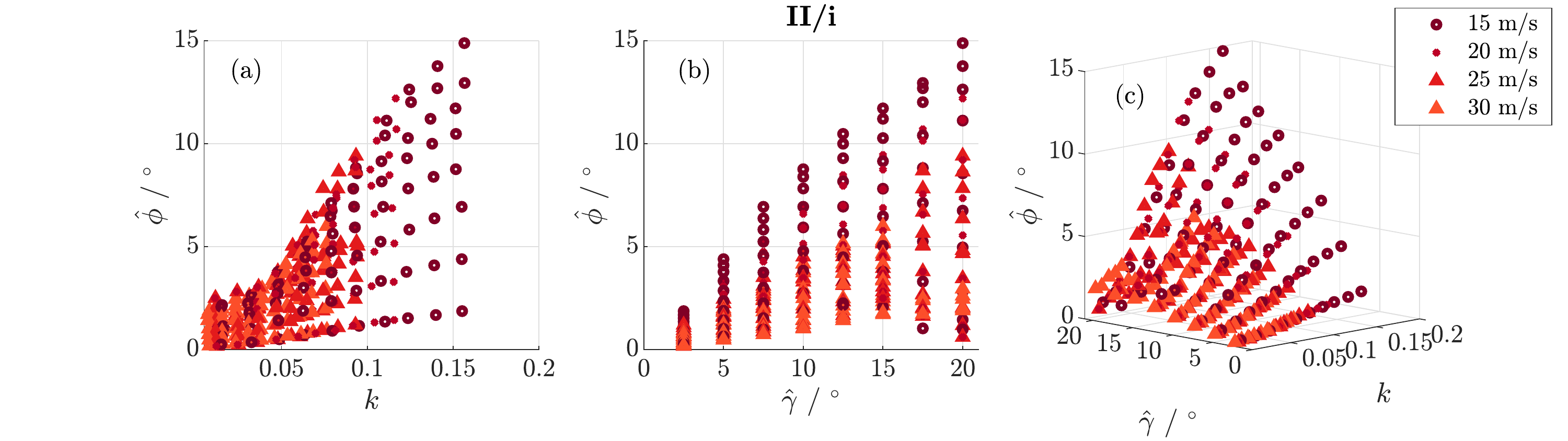}
\caption[Nine NACA all parameters]{Resulting AoI amplitudes for all measurements done with setup \textbf{II/i}. (a) Depending on the reduced frequency $k$. (b) Depending on the shaft amplitude $\hat{\gamma}$. (c) 3D view of results.}
\label{fig:Nine_NACA_all}
\end{figure*}

\begin{figure*}[h]
\centering
\includegraphics[width=0.99\linewidth]{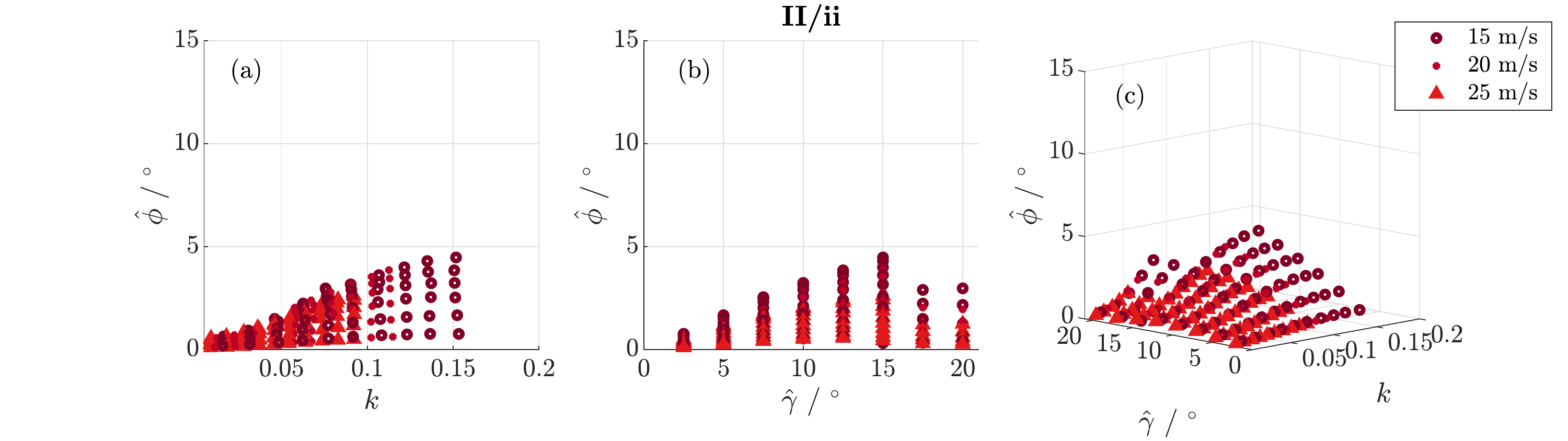}
\caption[Six NACA all parameters]{Resulting AoI amplitudes for all measurements done with setup \textbf{II/ii}. (a) Depending on the reduced frequency $k$. (b) Depending on the shaft amplitude $\hat{\gamma}$. (c) 3D view of results.}
\label{fig:Six_NACA_all}
\end{figure*}

\begin{figure*}[h]
\centering
\includegraphics[width=0.99\linewidth]{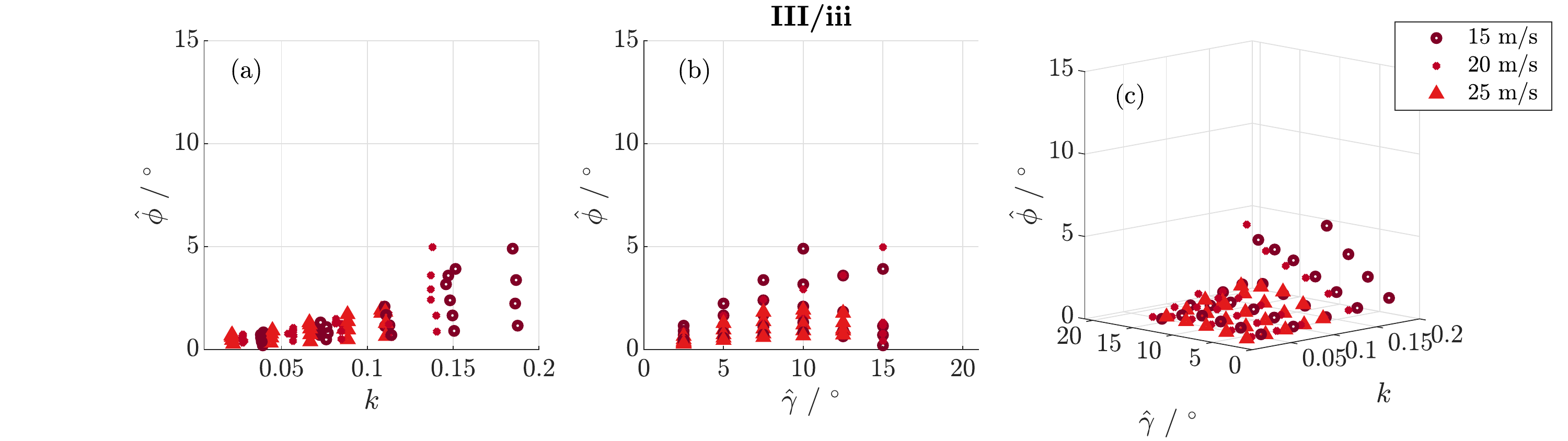}
\caption[Two NACA all parameters]{Resulting AoI amplitudes for all measurements done with setup \textbf{III/iii}. (a) Depending on the reduced frequency $k$. (b) Depending on the shaft amplitude $\hat{\gamma}$. (c) 3D view of results.}
\label{fig:Two_NACA_all}
\end{figure*}
\newpage

\section*{Velocity Modulation}
\begin{figure*}[h]
\centering
\includegraphics[width=0.99\linewidth]{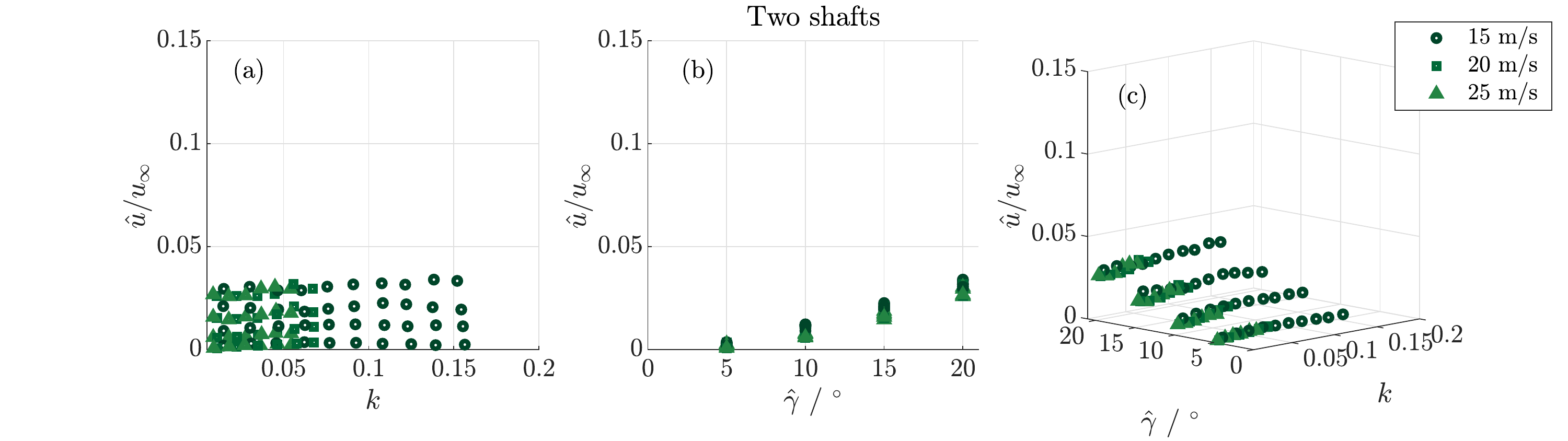}
\caption[Velocity fluctuation two shafts]{Resulting velocity variation amplitudes for all measurements using two shafts. (a) Depending on the reduced frequency $k$. (b) Depending on the shaft amplitude $\hat{\gamma}$. (c) 3D view of results.}
\label{fig:Vel_Fluc_Two_all}
\end{figure*}

\begin{figure*}[h]
\centering
\includegraphics[width=0.99\linewidth]{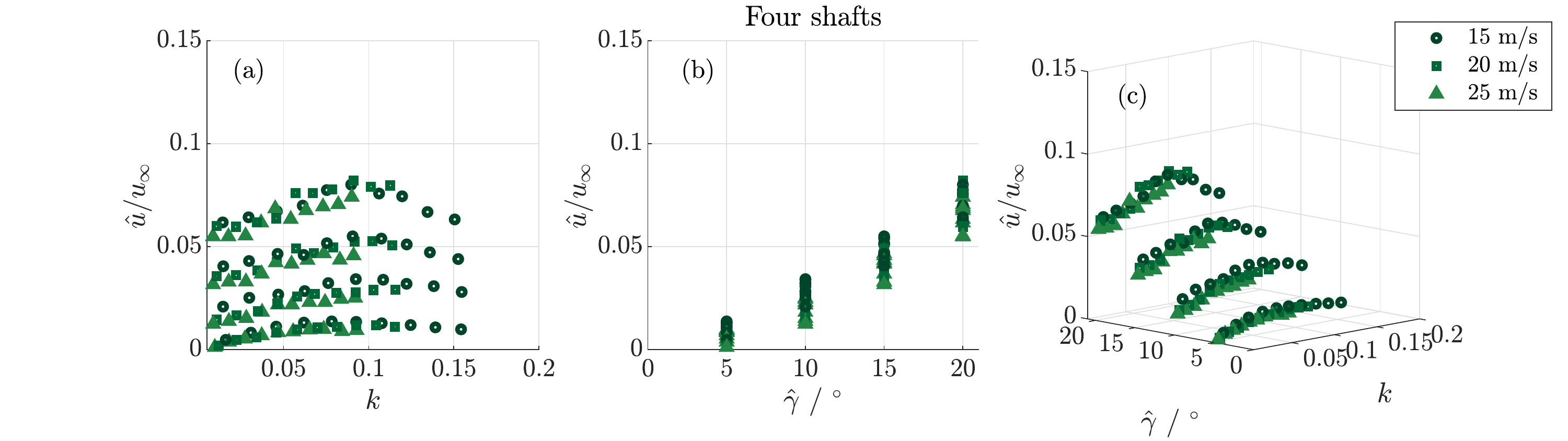}
\caption[Velocity fluctuation two shafts]{Resulting velocity variation amplitudes for all measurements using four shafts. (a) Depending on the reduced frequency $k$. (b) Depending on the shaft amplitude $\hat{\gamma}$. (c) 3D view of results.}
\label{fig:Vel_Fluc_Four_all}
\end{figure*}